\documentclass[a4paper,11pt]{article}

\title{Analysis of Lepton Flavor Violating \\ $\tau^\pm \to \mu^\pm \mu^\pm \mu^\mp$  Decays }

\author{\hspace{-3em} A.~Matsuzaki$^{1}$\footnote{akihiro@eken.phys.nagoya-u.ac.jp}
 ~and A.~I.~ Sanda$^{2}$
\footnote{sanda@kanagawa-u.ac.jp}\\[4mm]
\hspace{-3em}{ \small\textit{
1) Department of Physics, Nagoya University, Chikusa-Ku Furo-cho
    Nagoya 464-8602 Japan, }}\\
  \hspace{-3em} { \small\textit{  2)  Faculty of Engineering, Kanagawa University, Yokohama 221-8686, Japan  }}
}

\usepackage{bm}
 \def \gse {\hspace{0.3em}\raisebox{0.4ex}{$>$}\hspace{-0.75em}\raisebox{-.7ex}{$\sim$}\hspace{0.3em}}

\def \lse {\hspace{0.3em}\raisebox{0.4ex}{$<$}\hspace{-0.75em}\raisebox{-.7ex}{$\sim$}\hspace{0.3em}}

\def \in #1 #2 {\int \limits_{#1}^{#2}}
\def\Lag{\mathcal{L}} 

\usepackage[dvips]{graphicx}
\usepackage{amssymb}
\usepackage{amsmath}
\usepackage{graphicx} 
\usepackage{color}
\usepackage{axodraw}
\usepackage{cases}

\usepackage{amsmath}

\usepackage{pstricks}

\newcommand{\localinput}[1]{{
  \renewcommand{\documentclass}[2][dummy]{}
  \renewcommand{\usepackage}[2][dummy]{}
  \renewenvironment{document}{}{}
  \def\jobname{#1}
  \input{#1}
}}
\newcommand{\sla}[1]{\ooalign{\hfil\hspace{-0.1ex}\raise.2ex\hbox{$\not \phantom{#1}$}\hfil\crcr  $#1$}}

\begin{document}

\maketitle


\begin{abstract}
    Supposing only Lorentz and the gauge invariances of the Lagrangian, we derive energy and angular distributions for $\tau^\pm \to \mu^\pm \mu^\pm \mu^\mp$ lepton flavor violating decay process.    
  Using these results, we discuss methods to determine the parameters associated with the lepton flavor violating interactions.
  \end{abstract}

\newpage

\section{Introduction}

    Lepton flavor violating (LFV) decays, for example, $\mu \to e \gamma$, $\tau \to \mu \gamma$, $ \mu $-$e$ conversion and $ \tau \to 3 \mu $ are being studied  extensively by experimentists \cite{tau mu gamma}-\cite{belle tau3mu}, and by theorists \cite{Okada}-\cite{mu e conversion}. 
    Especially, a large number of $\tau$ decay events are collected in B Factories and we hope that LFV decay mode of $\tau$ may be found in the near future.  
   In that case, the Super B Factory \cite{superB}  might give us the large number of $ \tau \to 3 \mu $ events. 
   We might also hope to observe the energy and polarization distributions.

   In $ \tau \to  3\mu$ decay, we can measure all the energies and directions of the final state muons.
       For definiteness, let's say that we want to investigate the $\tau^+$ decay.
      B Factories generate back to back $\tau^+ \tau^-$ pairs. 
     As mentioned in Section \ref{sec full}, the polarization of $\tau^+$ can be observed statistically by correlation of momenta of decay products of both $\tau^+$ and $\tau^-$ \cite{tautau}.
      For $ \mu^+ \to  e^+ e^+ e^-$  process, the differential  branching ratio which is a function of energies of 2 positrons in the final state and the polarization of $\mu^+ $ in initial state is already derived in Ref. \cite{Okada}. 
The structure of weak interaction which causes the $\mu\to e\nu\bar{\nu}$ decay was investigated by Michel \cite{michel parameter}.
He introduced Michel parameters which proved to be very useful.
 We follow the similar strategies and formulate the method with which we can probe the structure of LFV interactions starting from a general Lagrangian.

\begin{figure} 
\begin{minipage}[t]{0.47\textwidth}
   \includegraphics[keepaspectratio=true,height=25mm]{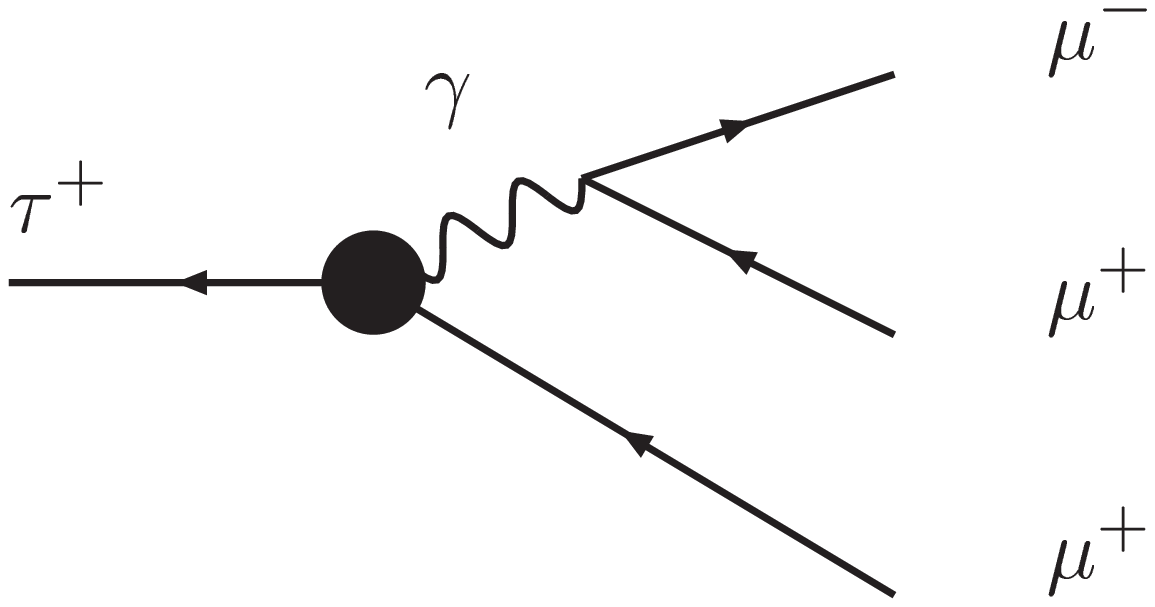}
   \end{minipage}
 \hfill
 \begin{minipage}[t]{0.47\textwidth}
   \includegraphics[keepaspectratio=true,height=25mm]{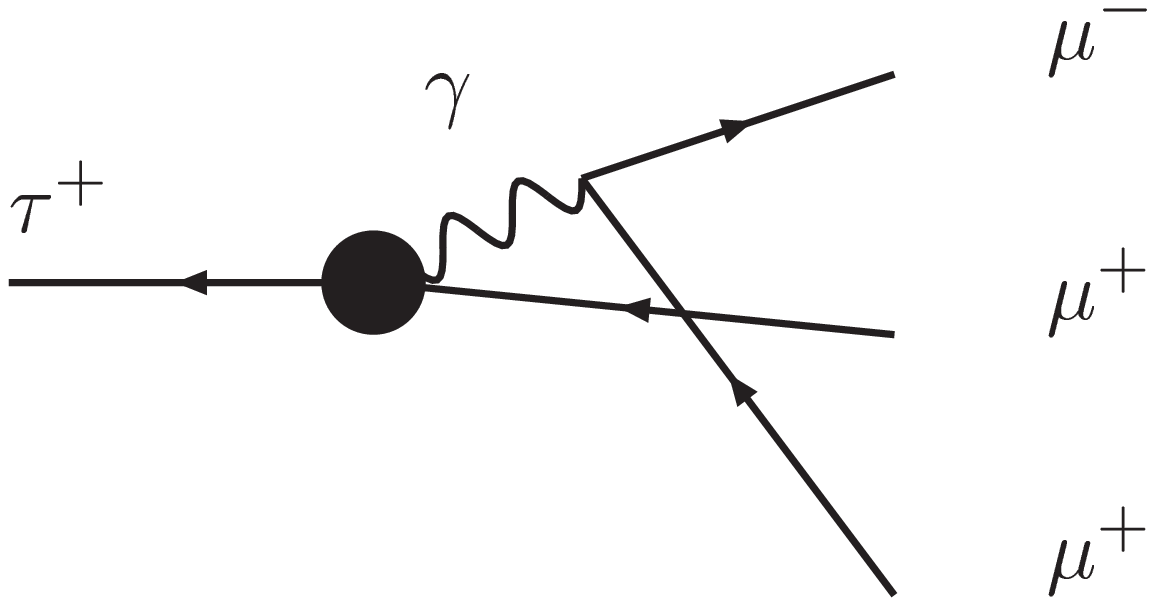}
   \end{minipage}
  \label{fig:photo}
  \caption{These are the diagrams produced from $\mathcal{L}_\gamma$.}
\end{figure}

The general Lagrangian for $\tau^+$ decaying to $\mu^+\mu^+\mu^-$ supposing Lorentz invariance  and gauge invariance  is written as:


\begin{align} \begin{split}\label{lag}
\Lag=\Lag_{FF}+\Lag_{\gamma},
\end{split} \end{align}

\begin{align} \begin{split}\label{lag4f} 
&\Lag_{FF}=- 2\sqrt{2}G_F \Bigr\{
                \hspace{1em}        g_{1}(\bar{\tau}_R \mu_L)(\bar{\mu}_R  \mu_L)
            +g_{2}(\bar{\tau}_L \mu_{R})(\bar{\mu}_L  \mu_R)  \\
 &   \hspace{8em}     +g_{3}(\bar{\tau}_R \gamma_\alpha \mu_R)(\bar{\mu}_R \gamma^\alpha  \mu_R)
            +g_{4}(\bar{\tau}_L \gamma_\alpha \mu_L)(\bar{\mu}_L \gamma^\alpha  \mu_L)   \\
 &   \hspace{8em}   +g_{5}(\bar{\tau}_R \gamma_\alpha \mu_R)(\bar{\mu}_L \gamma^\alpha  \mu_L)
            +g_{6}(\bar{\tau}_L \gamma_\alpha \mu_L)(\bar{\mu}_R \gamma^\alpha  \mu_R) \Bigl\},    
            \end{split} \end{align}
            
\begin{align} \begin{split}
\Lag_{\gamma}=&- 2\sqrt{2}G_F m_\tau\Bigr\{
              A_R\bar{\tau}_R\sigma^{\alpha\beta}\mu_L F_{\alpha\beta}
            + A_L\bar{\tau}_L\sigma^{\alpha\beta}\mu_R F_{\alpha\beta}\Bigl\}  \\
         &+\bar{\mu} (i D^\alpha\gamma_\alpha -m_\mu)  \mu         
              -\frac{1}{4}   F^{\alpha\beta}F_{\alpha\beta},
    \end{split} \end{align}
      where  $m_\mu$ and $m_\tau$ are the masses of the $\mu^\pm$ and $\tau^\pm$, respectively, $G_F $ is the Fermi constant. $ \{ \bar{\tau}_L, \bar{\mu}_L,  \mu_R\}$ and $\{\bar{\tau}_R, \bar{\mu}_R , \mu_L\}$   are the Dirac spinors with the helicity operators, $(1\pm \gamma_5)/2$, respectively.
       For example,  $\bar{\tau}_R=\bar{\tau}(1 - \gamma_5)/2$ and $ \mu_R =(1 + \gamma_5) \mu /2$.  
       $\sigma^{\alpha\beta}=\frac{i}{2}(\gamma^\alpha\gamma^\beta-\gamma^\beta\gamma^\alpha)$, 
      $D^\alpha=\partial^\alpha+ieA^\alpha$, $F^{\alpha \beta}=\partial^\alpha A^\beta-\partial^\beta A^\alpha$, $A^\alpha$ is the photon field and $e=-|e|$ is the electron charge. 
      $\Lag_{\gamma}$ represents $\tau \to \mu \gamma$ interaction as well as the terms present in ordinary QED. These terms generate diagrams in Fig. \ref{fig:photo}.
     So, $A_L$ and $A_R$ are the coefficients of interactions in which the intermediate photon has the left polarization, and the right polarization, respectively.
    The coefficients $g_1\sim g_6$ in $\Lag_{FF}$ are the coefficients of various 4-Fermi type interactions. 
    These terms generate diagrams in Fig. \ref{fig:4fermi}.
  Using the Fierz transformation, it is shown in the Appendix A that Eq. (\ref{lag}) is the general Lagrangian.

   \begin{figure}
\begin{minipage}[t]{0.47\textwidth}
   \includegraphics[keepaspectratio=true,height=25mm]{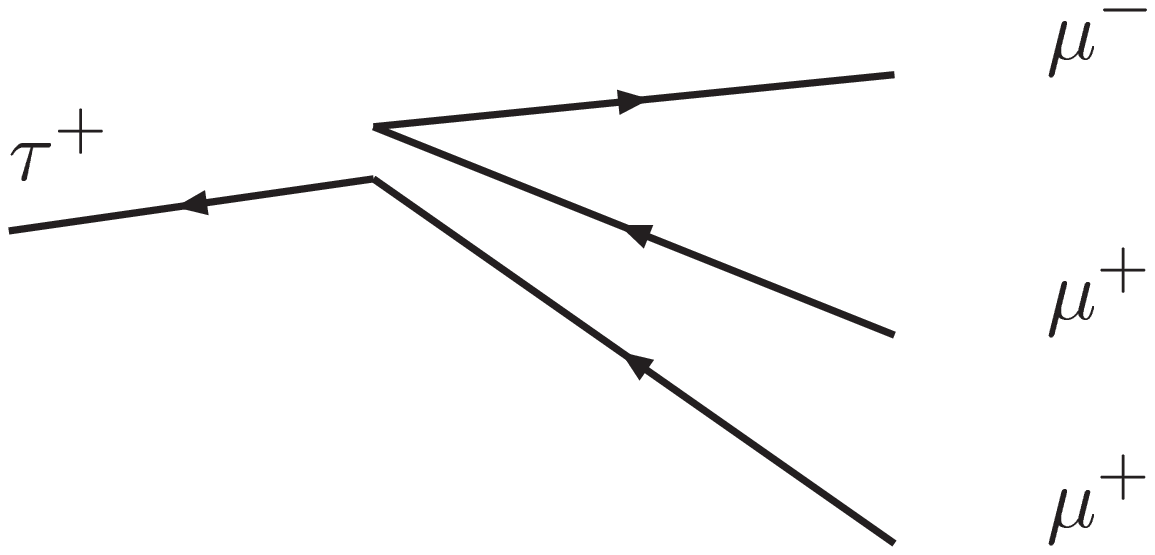}
 \end{minipage} 
 \hfill
 \begin{minipage}[t]{0.47\textwidth}
   \includegraphics[keepaspectratio=true,height=25mm]{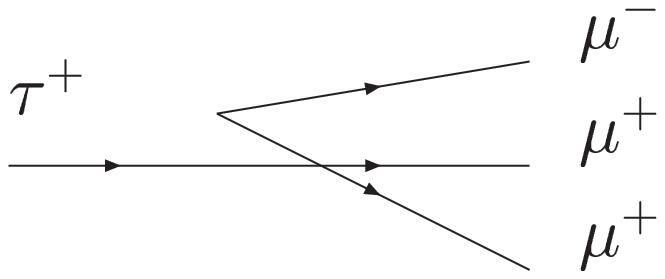}
 \end{minipage}
 \caption{These are the diagrams produced from $\mathcal{L}_{FF}$.}
  \label{fig:4fermi}
  \end{figure}

 For convenience, we summarize our major results, here assuming that enough events are collected. 
 \begin{enumerate}
  \item We have shown that  the absolute values of the coefficients  $|g_1|^2/16+|g_3|^2+|g_2|^2/16+|g_4|^2$, $|g_5|^2+|g_6|^2$, $|A_R|^2+|A_L|^2$, $Re[g_3 A_L^*]+Re[g_4 A_R^*]$ and $Re[g_5 A_L^*]+Re[g_6 A_R^*]$ can be obtained by measuring the energy distributions of decay products of $\tau^+$.
  \item We have shown that  the absolute values of the coefficients  $|g_1|^2/16+|g_3|^2$, $|g_2|^2/16+|g_4|^2$, $|g_5|$, $|g_6|$, $|A_R|$, $|A_L|$, $Re[g_3 A_L^*]$, $Re[g_4 A_R^*]$, $Re[g_5 A_L^*]$, $Re[g_6 A_R^*]$, $Im[g_3 A_L^*]+Im[g_4 A_R^*]$ and $Im[g_5 A_L^*]+Im[g_6 A_R^*]$ can be obtained by measuring the momentum distribution of decay products of $\tau^-$ in addition to the energy distributions of decay products of $\tau^+$.
  \item In some suitable cases, we can determine $|g_1|$, $|g_2|$, $|g_3|$ and $|g_4|$, independently, and phases $\arg[g_4 A_R^*]$, $\arg[g_3 A_L^*]$, $\arg[g_6 A_R^*]$ and  $\arg[g_5 A_L^*]$ as shown in section \ref{physics implication}.
  \item It is interesting to note that there is a possibility that some information on  $\tau\to \mu\gamma$ could be obtained before  $\tau\to \mu\gamma$ decay is measured.  For example, $Re[g_4 A_R^*]$ and $Re[g_3 A_L^*]$ arise from the interference terms between the four Fermi interaction amplitudes and the $\tau\to \mu\gamma$ decay amplitudes.
The decay rate for $\tau\to \mu\gamma$ is quadratic in  $|A_R|$ or $|A_L|$ while the observables given in  $Re[g_4 A_R^*]$ and $Re[g_3 A_L^*]$ are linear in  $|A_R|$ and $|A_L|$.
   So the interference effect may be seen before the decay rate for $\tau\to \mu\gamma$ is seen.
\end{enumerate}

  This paper is organized as follows.  Section \ref{sec full} gives the differential branching ratio of $ \tau^+ \to  \mu^+  \mu^+ \mu^-$ decay.  
   Also, a general formula which yields information on the current structure for $ \tau^+ \to  \mu^+  \mu^+ \mu^-$ decay is derived. 
   In section \ref{sec 3}, we pick energetic one of the two $\mu^+$ n the final state and analyze its energy dependence. 
    In section \ref{phys of 1}, we discuss physics implications of our results using only the results of section \ref{sec 3}.
 In section \ref{Energymu3},  we give the energy dependence of  $\mu^-$.
  In section \ref{phys of 3}, we discuss physics implications of our results using only the results of section \ref{Energymu3}.
  In section \ref{E 1,3}, we discuss physics implications of our results using the results of sections  \ref{phys of 1} and \ref{phys of 3}.
    In section \ref{sec 5}, we give  $\tau^+$ polarization dependence of the branching ratio in addition to the energetic $\mu^+$ energy dependence and analyze the results. 
    In section \ref{physics implication}, we discuss physics implications of our results using the results derived until previous section.
   Section \ref{conclud} contains the concluding remarks.


\section{General Formula}\label{sec full}
In this section, we derive the differential branching ratio for $ \tau^+ \to  \mu^+  \mu^+ \mu^-$  decay including $\tau^+$ polarization, and the general formula for the observables which are relevant for the actual experimental situation.

   The final state contains two $\mu^+$ mesons.
       The one with higher energy is denoted as $\mu_1$. 
    The other is denoted  as $\mu_2$.
      $\mu^-$ is denoted as $\mu_3$.

In the real experiment, the processes we want to detect are
\begin{align} \begin{split}\label{full-p}
e^+ e^- &\to  \tau^+(s^+) \  \tau^-(s^-)  \to  \nu_\tau+a+\mathrm{anything}  \\
    &\hspace{2em}\raisebox{1.44ex}{$_\lfloor$} \hspace{-1ex} \to    \mu_1  \ \mu_2 \  \mu_3 
\end{split} \end{align}
where $a$ is a particle which has the charge $-1$. So we must calculate the differential cross section for these processes.

Here, we use the center of mass frame of $ e^+ e^-$ initial state, which we call frame 1.
    In the rest frame of $\tau^+$, which we name frame 3, the momenta of $\mu_1$, $\mu_2$,  $\mu_3$ and $\tau^+$ are denoted as $p_1=(E_1,\mathbf{p}_1 )$, $p_2=(E_2,\mathbf{p}_2)$,  $p_3=(E_3,\mathbf{p}_3 )$ and   $p_\tau=(m_\tau,\mathbf{0})$, and finally the polarization of $\tau^+$ is denoted as $s^+$. 
  $s^-$ denotes the polarization of $\tau^-$ in the rest frame of $\tau^-$.
         Here, $ s^+ \cdot p_\tau  = s^- \cdot k_\tau  =0$ and $(s^\pm)^2=-1$ where $k_\tau$ is the momentum of $\tau^-$ in the $\tau^-$ rest frame which we call frame 2.

\begin{figure}[htbp]
  \begin{center}
    \includegraphics[ width=10cm]{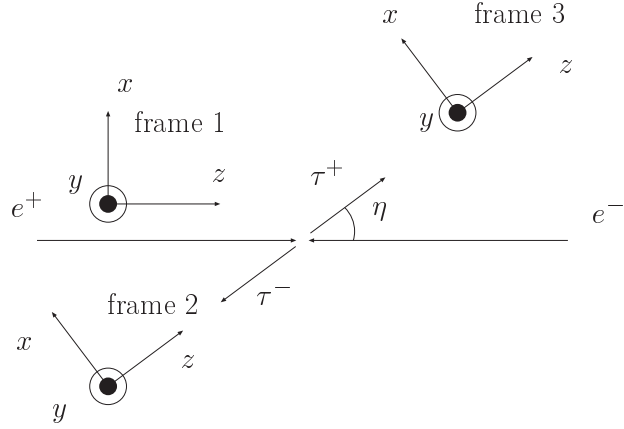}
  \end{center}
  \caption{The relation between frame 1, frame 2 and frame 3. $\eta$ is the angle between $\mathbf{p}_{e^+}'$ and $\mathbf{p}_\tau'$.}
    \label{fig:frame123.eps}
\end{figure}

For the  definiteness, we set the relations between frame 1, 2, 3, as follows and they are depicted in Fig. \ref{fig:frame123.eps}.
\begin{description}
  \item Frame 1 is the center of mass frame of $e^+ e^-$.
  Defining the $\tau^+$  momentum as  $\mathbf{p}_\tau'$  and the initial state positron momentum as $\mathbf{p}_{e^+}'$, we set the $z$ direction in this frame as the direction of $\mathbf{p}_{e^+}'$, $y$ direction is the same as $ \mathbf{p}_{e^+}' \times \mathbf{p}_{\tau}'$ and the $x$ direction as $(\mathbf{p}_{e^+}' \times \mathbf{p}_{\tau}')\times \mathbf{p}_{e^+}'  $.
 
  \item  Frame 2 is the $\tau^-$ rest frame. We set  the $z$ direction in this frame as the direction of $\mathbf{p}_{\tau}'$.  The $y$ direction is the same as that of the frame 1. The $x$ direction id defined by $(\mathbf{p}_{e^+}' \times \mathbf{p}_{\tau}')\times \mathbf{p}_{\tau}'  $.

  \item  Frame 3 is the $\tau^+$ rest frame, We set the directions in this frame are  the same as that of the frame 2. 

\end{description}

There are four ingredients that are used to compute the differential cross section for the process shown in Eq. (\ref{full-p}).
\begin{enumerate}
  \item The narrow width approximation where we approximate
  \begin{align} \begin{split}
e^+ e^- &\to  \tau^+(s^+) \  \tau^-(s^-)  \to  \nu_\tau+a+\mathrm{anything} \nonumber \\
    &\hspace{2em}\raisebox{1.44ex}{$_\lfloor$} \hspace{-1ex} \to    \mu_1  \ \mu_2 \  \mu_3 
\end{split} \end{align}
is described in appendix \ref{narrow width app}.
\item The differential cross section for $e^+ e^- \to \gamma^* \to \tau^+(s^+) \tau^-(s^-)$ is presented in appendix \ref{sec e+e-to taupair}.
  \item The differential branching ratio for $\tau^-(s^-) \to \nu_\tau+a+\mathrm{anything}$  is presented in appendix \ref{sec tau^-decay}.
  \item The differential branching ratio for $\tau^+(s^+)\to \mu_1 \mu_2 \mu_3$ is given below.
 \end{enumerate}
  The differential cross section for this process $d\sigma$ is written as \cite{Sanda}

\begin{align}\begin{split} \label{full-p-sigma}
 &\hspace{-2em}\frac{d\sigma}
{d\Omega \  dx_1 dx_2 d\Omega_\tau d\psi \ d^3 k_a }  \\
=&
\raisebox{-1ex}{\text{{\huge \textsf{S}}}}\hspace{-2em}   \raisebox{-3ex}{$\scriptstyle {s^+ , s^-}$}     \hspace{ 1em}
 \sum _{\pm s^+ ,  \pm s^-}\frac{d\sigma\bigl(e^+ e^- \to \tau^+(s^+) \tau^-(s^-)\bigr)}{d\Omega} \\ &\hspace{4em}\times
\frac{dBr\bigl(\tau^-(s^-) \to \nu_\tau+a+\mathrm{anything}\bigr)}{d^3 k_a}
 \\ &\hspace{4em} \times
 \frac{dBr\bigl(\tau^+(s^+)\to \mu_1 \mu_2 \mu_3 \bigr)}{dx_1 dx_2 d\Omega_\tau d\psi }.
\end{split} \end{align}
\textsf{S} implies sum over polarizations. 
$k_a$ is the momentum of the particle $a$ in $\tau^-$ rest frame.
The definitions of $\Omega_\tau$ and $\psi$ are written in next paragraph.

\begin{figure}[htbp]
  \begin{center}
    \includegraphics[keepaspectratio=true,height=40mm]{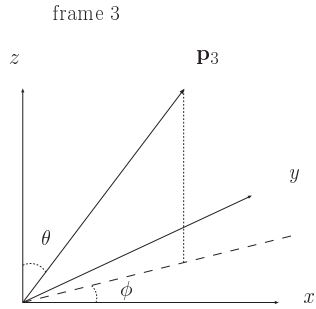}
  \end{center}
  \caption{Definition of $\theta$ and $\phi$ in frame 3}
  \label{fig:frame3thetaphi.eps}
\end{figure}

\begin{figure}[htbp]
  \begin{center}
    \includegraphics[keepaspectratio=true,height=50mm]{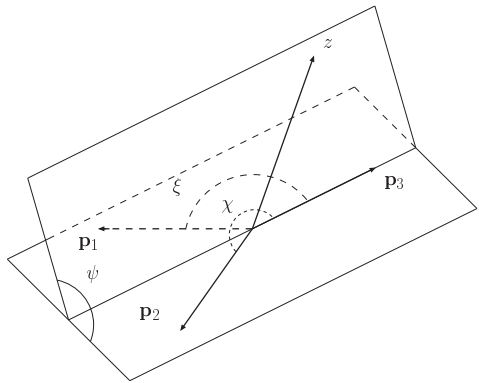}
  \end{center}
  \caption{$\psi$ is the angle between $\mathbf{p}_2$-$\mathbf{p}_3$ plane and $\mathbf{p}_3$-$z$ plane. $\xi$ is the angle between $\mathbf{p}_3$ and $\mathbf{p}_1$.  $\chi$ is the angle between $\mathbf{p}_3$ and $\mathbf{p}_2$. $0\le \xi\le \pi$.  $\pi\le \chi\le 2\pi$.}
  \label{fig:frame3p1p2.eps}
\end{figure}

      We have computed the branching ratio for $ \tau^+(s^+) \to  \mu_1 \mu_2 \mu_3$ including $m_\mu$ dependences. 
      However, for now, we confine our discussion where we can approximate $m_\mu/m_\tau=0$.
%
        The result is
\begin{align} \begin{split}\label{tau^+ full}
 \frac{dBr\bigl(\tau^+(s^+)\to \mu_1 \mu_2 \mu_3 \bigr)}{dx_1 dx_2 d\Omega_\tau d\psi }\hspace{-3em}&
 \\ =\frac{3}{2\pi^2}Br(\tau \to  \mu \nu \bar\nu)
  \Bigl[ & G_0(x_1,x_2)+ \sum_i \mathbf{s}^+ \cdot \mathbf{P}_i G_i^s(x_1,x_2) \Bigr], 
\end{split} \end{align}
   where the definitions of $G_0(x_1,x_2)$ and  $G_i^s(x_1,x_2)$ are written in  Appendix \ref{Appendix C}. 
$\mathrm{Br}(\tau\to \mu \nu\bar\nu)$ is the branching ratio of $\tau^+\to \mu^+ \nu_\mu \bar\nu_\tau$ decay. 
\begin{eqnarray}
x_1=\frac{2E_1}{ m_\tau},\ \ x_2=\frac{2E_2}{ m_\tau},\ \ x_3=\frac{2E_3}{ m_\tau}.
\end{eqnarray}
Note that $x_1$, $x_2$ and $x_3$ take values between 0 and 1, and $x_1+x_2+x_3=2$.
$\Omega_\tau$ is the solid angle defined in Fig. \ref{fig:frame3thetaphi.eps} and $\psi$ is the angle between the $\mathbf{p}_2$-$\mathbf{p}_3$ plane and $z$-$\mathbf{p}_3$ plane as defined in Fig. \ref{fig:frame3p1p2.eps}. 
$\mathbf{P}_i =\{ \hat{\mathbf{p}}_1 ,\ \hat{\mathbf{p}}_2 ,\ \hat{\mathbf{p}}_1 \times \hat{\mathbf{p}}_2  \}$ where $\hat{\mathbf{p}}_1=\mathbf{p}_1/ |\mathbf{p}_1|$ and $\ \hat{\mathbf{p}}_2=\mathbf{p}_2/ |\mathbf{p}_2|$.


Substituting the concrete representations,
 Eq. (\ref{full-p-sigma}) becomes  
\begin{align} \begin{split}\label{substituted }
& \frac{d\sigma}
{d\Omega \  dx_1 dx_2 d\Omega_\tau  d\psi \ d^3 k_a }    \\
&=\raisebox{-1ex}{\text{{\huge \textsf{S}}}}\hspace{-1.3em}   \raisebox{-3ex}{$\scriptstyle s^+ , s^-$}     \hspace{ 0.41em}
\sum _{\pm s^+ ,  \pm s^+}\frac{  \alpha^2 \beta }{4 q^2}\biggl[(1+\cos^2\eta +\frac{\sin ^2 \eta}{\gamma^2}) +(1+\cos^2\eta -\frac{\sin ^2 \eta}{\gamma^2}) s_z^+ s_z^- 
   \\
&\hspace{ 9.4em}-\beta^2 \sin ^2 \eta \  s_y^+ s_y^-+(1+\frac{1}{\gamma^2})\sin^2\eta \ s_x^+ s_x^-
-\frac{\sin 2\eta}{\gamma}(s_z^+s_x^- + s_x^+  s_z^-)\biggr]  
\\ &\hspace{1em}\times \frac{3}{2\pi^2}Br(\tau \to \mu \nu \bar\nu)
   \Bigl[G_0(x_1,x_2)+ \sum_i \mathbf{s}^+ \cdot \mathbf{P}_i G_i^s(x_1,x_2) \Bigr]  
\\ &\hspace{1em}\times Br\bigl(\tau^- \to \nu_\tau+a+\mathrm{anything}\bigr) \frac{2}{\pi m_\tau^3 \lambda_a}\left[ G_1^a(y_a)-  \mathbf{s}^- \cdot  \hat {\mathbf{k}}_a   G_2^a(y_a) \right].
\end{split} \end{align}

Carrying out the summation for $\pm s^+ $ and $ \ \pm s^-$ using $s^\pm_i s^\pm_j=\delta_{ij}$,  Eq. (\ref{substituted }) becomes
\begin{align} \begin{split}\label{sum pm s ^pm}
&\frac{d\sigma}
{d\Omega \  dx_1 dx_2 d\Omega_\tau  d\psi \ d^3 k_a }   
 \\&=
 4  Br(\tau \to \mu \nu \bar\nu) Br\bigl(\tau^- \to \nu_\tau+a+\mathrm{anything}\bigr)
\frac{3 \alpha^2 \beta }{4\pi^3 q^2 m_\tau^3 \lambda_a}   
\\ &\hspace{1em}\times
\biggl[ G_0(x_1,x_2) G_1^a(y_a)\bigl(1+\cos^2\eta +\frac{\sin ^2 \eta}{\gamma^2}\bigr)  
\\  &\hspace{2.5em} -\sum_i G_i^s(x_1,x_2)G_2^a(y_a)
\Bigl\{ \bigl(1+\cos^2\eta -\frac{\sin ^2 \eta}{\gamma^2}\bigr) \hat  k_{az}  P_{iz}-\beta^2 \sin^2\eta \  \hat k_{ay} P_{iy}  
\\  &\hspace{13.5em}+\bigl(1+\frac{1}{\gamma^2}\bigr) \sin^2\eta \  \hat k_{ax}  P_{ix} - \frac{\sin 2 \eta}{\gamma} (\hat  k_{ax}  P_{iz}+\hat  k_{az}  P_{ix})
\Bigr\}
\biggr].
\end{split} \end{align}
This expression is the general formula of the process (\ref{full-p}) as long as we stay sufficiently away from singularity at $m_\mu=0$.
In following sections, we'll start analyzing from this expression. 



\section{$G_0(x_1)$: Energy Dependence of $\mu_1$ }\label{sec 3}

   In this section, we derive the formulae convenient for investigating the structure of the LFV four Fermi interactions, using the $\mu_1$ energy dependence of the differential branching ratio for $\tau \to 3 \mu $ decay mode.

 Our first priority is to discuss the observable which is  easier to detect and analyze.
So here, we integrate the polarization dependence as follows.
First, we integrate over $d\Omega   d\Omega_\tau  d\psi$,
\begin{align} \begin{split}
&\frac{d\sigma}
{       dx_1 dx_2   \ d^3 k_a}     \\\hspace{-3em}
&=
   Br(\tau \to \mu \nu \bar\nu) Br\bigl(\tau^- \to \nu_\tau+a +\mathrm{anything}\bigr)
\frac{64 \alpha^2 \beta }{  q^2 m_\tau^3 \lambda_a}  
 G_0(x_1,x_2) G_1^a(y_a)(2  +\frac{1   }{ \gamma^2}).
\end{split} \end{align}
Next, we integrate over $d^3 k_a$,
\begin{align} \begin{split}\label{sigma-non-pol}
&\frac{d\sigma}
{       dx_1 dx_2   }  
  \\\hspace{-3em}
&=
 \frac{32 \pi \alpha^2 \beta }{  q^2   } (2  +\frac{1   }{ \gamma^2}) Br(\tau \to \mu \nu \bar\nu) Br\bigl(\tau^- \to \nu_\tau+a +\mathrm{anything}\bigr)
  \\ &\hspace{2em} \times 
    G_0 (x_1,x_2) . 
\end{split} \end{align}
  Eq. (\ref{sigma-non-pol}) allow us to obtain $G_0(x_1,x_2)$.
It has the arguments $x_1$ and $x_2$. 
    Even expression, however, is pretty complicated for a discussion here.
We thus discuss only the $x_1$ dependence integrating over $x_2$.
    For the physical region, we found that the effect of neglecting muon masses in the differential branching ratio for the $\tau \to 3 \mu$ decay introduces an error of $\mathcal{O}(2m_\mu/m_\tau)$.

    Now we define
\begin{align}\begin{split}\label{Eq +}
a_+&=\bigl(\frac{|g_1|^2}{16}+|g_3|^2 \bigr)+\bigl( \frac{|g_2|^2}{16}+|g_4|^2\bigr) \\
b_+&=|g_5|^2+|g_6|^2 \\
c_+&= |eA_R|^2 + |eA_L|^2 \\
d_+&= -\bigl( Re[g_3 e A_L^*]  +Re[g_4 e A_R^*]\bigr)  \\
e_+&= -\bigl(Re[g_6 e A_R^*]  + Re[g_5 e A_L^*]\bigr).  
\end{split}\end{align}

 Then, $G_0$ integrated over $x_2$ is  
\begin{align} \begin{split}
 G_0(x_1)&=  \in {1-x_1} {x_1}  dx_2 \ G_0(x_1,x_2)   \\ 
&= 
    \frac{1}{3}  J_1 (2x_1-1)\Bigl\{ 6(1-x_1)  (2x_1-1) 
   \\&\hspace*{7.5em}  +\rho_{1a}  (6x_1-5 )(2x_1-1)   
      \\&\hspace*{7.5em}   +\frac{8}{3}\rho_{1b} (3x_1-2)(x_1-1)   
 \Bigr\} 
 \\&+ \frac{4}{3} c_+ \Bigl\{\frac{2(2x_1-1)(x_1^2-x_1+1)}{ (1-x_1)} +3 (2x_1^2-2x_1+1)\log\bigl[\frac{x_1}{1-x_1}\bigr]\Bigr\}, \\
\end{split} \end{align}

where

\begin{align} \begin{split}\label{J_1}
J_1&=\frac{1}{3}(10a_+ +7 b_+ + 72d_+ +54e_+ )
\\J_1 \rho_{1a}&=\frac{1}{2}( 4a_+ + b_+ +48d_+ +12e_+ ) 
\\J_1 \rho_{1b}&=\frac{9}{4} ( b_+ +8e_+ ).  
\end{split} \end{align}
  We have defined $ \rho_{1a}$ and $\rho_{1b}$  to take values between $0$ and $1$.   We now discuss how  $c_+$, $J_1$, $\rho_{1a}$ and $\rho_{1b}$ can be determined from the $x_1$ dependence of  $G_0$.   
  
  First, to determine $c_+$, it is convenient to define
   the function 
\begin{eqnarray}
F_1(x_1)=
\frac{3}{8}(1-x_1)  G_0(x_1).
 \end{eqnarray}
    By choosing the kinematics such that $x_1\to 1$ for the function $F_1(x_1)$, we can obtain  $c_+$ as
\begin{eqnarray}
F_1(x_1)\bigr|_{x_1=1}
=c_+  (2x_1-1)(x_1^2-x_1+1)\bigr|_{x_1=1} 
 = c_+.   
 \end{eqnarray}
 
     Next, we subtract the term containing the coefficient $c_+ $ and define another function,
\begin{align} \begin{split}
F_2(x_1) &= 
  \frac{ G_0(x_1) -(c_+,x_1\ \mathrm{term})}{(2x_1-1)}
\\&=
 \frac{1}{3} J_1 \Bigl\{ 6(1-x_1)  (2x_1-1) 
   \\&\hspace{3.4em}  +\rho_{1a}  (6x_1-5 )(2x_1-1) 
   \\&\hspace{3.4em}   +\frac{8}{3}\rho_{1b} (3x_1-2)(x_1-1)  
 \Bigr\},
 \end{split} \end{align} 
  where 
 \begin{align} \begin{split}
(c_+,x_1\ \mathrm{term})=\frac{4}{3}c_+ \Bigl\{\frac{2(2x_1-1)(x_1^2-x_1+1)}{ (1-x_1)} +3 (2x_1^2-2x_1+1)\log\bigl[\frac{x_1}{1-x_1}\bigr]\Bigr\}.
\end{split} \end{align}
     We can then determine the parameters $J_1$, $\rho_{1a}$ and $\rho_{1b}$ from the shape of $F_2(x_1)$ using 
\begin{eqnarray}
\int_{1/2}^1 F_2(x_1)dx_1=\frac{1}{12}J_1,
\end{eqnarray}
\begin{eqnarray}
F_2(x_1)\bigr|_{x_1=1}=\frac{1}{3}J_1\rho_{1a}
\end{eqnarray}
 and 
\begin{eqnarray}
F_2(x_1)\bigr|_{x_1=\frac{1}{2}}=\frac{2}{9} J_1 \rho_{1b},
\end{eqnarray}
 respectively.

\section{Physics from $G_0(x_1)$ Distribution of $\mu_1$}\label{phys of 1}

The conclusion of previous section is how to determine the parameters, $c_+$, $J_1$, $\rho_{1a}$, $\rho_{1b}$.
Using only these parameters, $a_+$, $b_+$, $d_+$ and $e_+$ can't be determined.
However, if we take some special cases which are explained bellow, we can restrict the allowed regions of $a_+$, $b_+$, $d_+$ and $e_+$.  

\subsection{What happens if one of $a_+$, $b_+$, $c_+$, $d_+$, $e_+$ $=0$}
If we impose that one of $a_+$, $b_+$, $c_+$, $d_+$ and $e_+$ is zero from other experimental results or supposing some specific models, the results are as follows.
\\ 
When $a_+=0$, also $d_+=0$ and 
\begin{align} \label{a=0,1}
J_1&=\frac{1}{3}(7b_++54e_+)       \nonumber\\
J_1\rho_{1a}&=\frac{1}{2}(b_++12e_+) \\
J_1\rho_{1b}&=\frac{9}{4}(b_++8e_+) .    \nonumber
 \end{align}
When $b_+=0$, also $e_+=0$ and  
\begin{align} \label{b=0,1}
J_1&=\frac{1}{3}(10a_++72d_+)      \nonumber\\
J_1\rho_{1a}&=\frac{1}{2}(4a_++48d_+) \\
J_1\rho_{1b}&=0  .    \nonumber
 \end{align}
When $c_+=0$, also $d_+=e_+=0$ and
\begin{align}
J_1&=\frac{1}{3}(10a_++7b_+)         \nonumber\\
J_1\rho_{1a}&=\frac{1}{2}(4a_++b_+)   \\
J_1\rho_{1b}&=\frac{9}{4}b_+        .    \nonumber
\end{align}
When $d_+=0$, 
\begin{align} \begin{split} 
J_1&=\frac{1}{3}(10a_+ +7 b_+  +54e_+ )
\\J_1 \rho_{1a}&=\frac{1}{2}( 4a_+ + b_+  +12e_+ ) 
\\J_1 \rho_{1b}&=\frac{9}{4} ( b_+ +8e_+ ).  
 \end{split} \end{align}
When $e_+=0$, 
\begin{align} \begin{split} 
J_1&=\frac{1}{3}(10a_+ +7 b_+ + 72d_+  )
\\J_1 \rho_{1a}&=\frac{1}{2}( 4a_+ + b_+ +48d_+  ) 
\\J_1 \rho_{1b}&=\frac{9}{4}  b_+.  
 \end{split} \end{align}
If any one of $a_+$, $b_+$, $c_+$, $d_+$ and $e_+$ is zero, we can determine $a_+$, $b_+$, $c_+$, $d_+$ and $e_+$ from only the $x_1$ distribution.

\subsection{Results if either $a_+\not=0$, $b_+\not=0$, $d_+\not=0$ or $e_+\not=0$
}\label{6.8}
Similarly,
When only $a_+\not=0$, 
\begin{align}
J_1&=\frac{10}{3}a_+   \nonumber\\
\rho_{1a}&=\frac{3}{5} \\
\rho_{1b}&=0.\nonumber
\end{align}
When only $b_+\not=0$, 
\begin{align}
J_1&=\frac{7}{3}b_+  \nonumber\\
\rho_{1a}&=\frac{3}{14} \\
\rho_{1b}&=\frac{27}{28}.\nonumber
\end{align}
When only $a_+,\ c_+,\ d_+\not=0$ but $a_+$ is negligible, 
\begin{align} \begin{split}
J_1&=24d_+
\\ \rho_{1a}&=1  
\\ \rho_{1b}&=0.  
\end{split} \end{align}
When only $b_+,\ c_+,\ e_+\not=0$ but $b_+$ is negligible, 
\begin{align} \begin{split}
J_1&=18e_+
\\ \rho_{1a}&=\frac{1}{3}  
\\ \rho_{1b}&=1.  
\end{split} \end{align}
In these case, $\rho_{1a}$ and $\rho_{1b}$ are independent from $a_+$, $b_+$, $d_+$ and $e_+$.

\subsection{What happens if $\tau\to\mu\gamma$ dominates
}\label{c>>,1}

If $\tau\to\mu\gamma$ type interaction which has coefficient $c_+$ is much larger than other interactions, only $c_+$, $d_+$ and $e_+$ may be determined since  $d_+$ and $e_+$ are not quadratic but linier of 4-fermi type interactions. 
In this case,
\begin{align} \begin{split}
J_1&=6(4d_++3e_+)\\
J_1\rho_{1a}&=6(4d_++e_+)\\
J_1\rho_{1b}&=18e_+.
\end{split} \end{align}
From Eqs. (\ref{Eq h}) and (\ref{Eq h2}) in appendix \ref{6.1}, 
\begin{align} \begin{split}
a_+&\ge\frac{d_+^2}{c_+}=\frac{1}{c_+}\left(\frac{J_1(1-\rho_{1b})}{24}\right)^2\\
b_+&\ge\frac{e_+^2}{c_+}=\frac{1}{c_+}\left(\frac{J_1\rho_{1b}}{18}\right)^2.
\end{split} \end{align}
So, we can determine  $a_+$ and $b_+$ lower limits, though we can't determine  $a_+$ and $b_+$, directly. 

\subsection{Lower limits of $c_+$ when $\tau\to\mu\gamma$ is highly suppressed}
We give three types of $c_+$ lower limits. 
Especially, these are very useful when $\tau\to\mu\gamma$ is highly suppressed compared with $\tau\to 3\mu$ and $c_+$ cannot determine directly.

First, we give the relations,
\begin{align} \begin{split}\label{29}
                       \frac{4}{9}J_1 \rho_{1b}&=b_++8e_+
\\3J_1- 5J_1\rho_{1a}-           2J_1 \rho_{1b}&=-12 (4 d_+ + e_+)
\\3J_1- 3J_1\rho_{1a}-\frac{22}{9}J_1 \rho_{1b}&=4(a_+-2e_+)
\\3J_1+  J_1\rho_{1a}-\frac{10}{3}J_1 \rho_{1b}&=12(a_++8d_+)
\\6J_1-10J_1\rho_{1a}- \frac{8}{3}J_1 \rho_{1b}&=3(b_+-32d_+)
\\3J_1- 3J_1\rho_{1a}-           2J_1 \rho_{1b}&=4a_++b_+
\end{split} \end{align}
 from Eqs. (\ref{J_1}).

Using Eqs. (\ref{29}), (\ref{Eq h}) and (\ref{Eq h2}),
\begin{align} \begin{split}
\frac{1}{12^2\times 5}
\frac{(3J_1- 5J_1\rho_{1a}-2J_1 \rho_{1b})^2}{3J_1- 3J_1\rho_{1a}-           2J_1 \rho_{1b}}
&=\frac{(4d_++e_+)^2}{5(4a_++b_+)}
\\&\le
\frac{(4d_++e_+)^2}{5\left(4\frac{d_+^2}{c_+}+\frac{e_+^2}{c_+}\right)}
\\&=c_+\frac{(4\frac{d_+}{e_+}+1)^2}{5\left(4\frac{d_+^2}{e_+^2}+1\right)}
\\&\le c_+.
\end{split} \end{align}
This becomes equality when $a_+ c_+-d_+^2=0$, $b_+ c_+-e_+^2=0$ and $d_+=e_+$.

When $a_+\gg c_+$, if $6J_1-10J_1\rho_{1a}- \frac{8}{3}J_1 \rho_{1b}=3(b_+-32d_+)<0$ and no special cancelation between $b_+$ and $32d_+$, then $b_+-32d_+=\mathcal{O}(-32d_+)$ and $b_+\le\mathcal{O}(32d_+)\ll 4a_+$.
So, using Eq. (\ref{Eq h}),
\begin{align} \begin{split}\label{denom a}
&\frac{1}{256\times 3^2}
\frac{(6J_1-10J_1\rho_{1a}-\frac{8}{3}J_1 \rho_{1b})^2}{3J_1- 3J_1\rho_{1a}- 2J_1 \rho_{1b}}
\\&\hspace{5em}=
\frac{(b_+-32d_+)^2}{256(4a_++b_+)}
\le
\mathcal{O}\left(\frac{32^2d_+^2}{256\times4a_+}\right)
\le
\mathcal{O}(c_+).
\end{split} \end{align}
We note here that $4a_++b_+$ is similar to $4( a_+-2e_+)$ and $4(a_++8d_+)$ in this situation. 
So, we have similar result if the denominator of left hand side of Eq. (\ref{denom a}),     $
  3J_1- 3J_1\rho_{1a}-           2J_1 \rho_{1b}=4a_++b_+
 $ is exchanged by $ 
  3J_1- 3J_1\rho_{1a}-\frac{22}{9}J_1 \rho_{1b}=4(a_+-2e_+)$ or $
(3J_1+  J_1\rho_{1a}-\frac{10}{3}J_1 \rho_{1b})/3=4(a+8d_+).$

Similarly, when $b_+\gg c_+$, if $3J_1- 3J_1\rho_{1a}-\frac{22}{9}J_1 \rho_{1b}=4(a_+-2e_+)<0$ and no special cancelation between $a_+$ and $2e_+$, then $a_+-2e_+=\mathcal{O}(-2e_+)$ and $a_+\le\mathcal{O}(2e_+)\ll b_+/4$.
So, using Eq. (\ref{Eq h2}),
\begin{align} \begin{split}\label{denom b}
&\frac{1}{8^2}
\frac{(3J_1- 3J_1\rho_{1a}-\frac{22}{9}J_1 \rho_{1b})^2}{3J_1- 3J_1\rho_{1a}- 2J_1 \rho_{1b}}
\\&\hspace{5em}=
\frac{(a_+-2e_+)^2}{4(4a_++b_+)}
\le
\mathcal{O}\left(\frac{2^2e_+^2}{4b_+}\right)
\le
\mathcal{O}(c_+).
\end{split} \end{align}
We note here that $a_++b_+/4$ is similar to $a_+-2e_+$ and $ a_++8d_+$ in this situation. 
So, we have similar result if the denominator of left hand side in Eq. (\ref{denom b}),    $
  3J_1- 3J_1\rho_{1a}-           2J_1 \rho_{1b}=4a_++b_+
  $ is exchanged by $
                \frac{4}{9}J_1 \rho_{1b}=b_++8e_+$ or $
(6J_1-10J_1\rho_{1a}- \frac{8}{3}J_1 \rho_{1b})/3=b_+-32d_+.$

If we suppose that $b_+=0$ from some special models or other experimental results, we have $c_+$ lower limit from (\ref{b=0,1}) and (\ref{Eq h}), 
\begin{align} \begin{split}
\frac{1}{1728}\frac{(3J_1 -5J_1\rho_{1a})^2}{J_1-J_1\rho_{1a}}
=\frac{d_+^2}{a_+}
\le
c_+.
\end{split} \end{align}

Similarly, from (\ref{a=0,1}) and (\ref{Eq h2}), if we suppose that $a_+=0$ from some special models or other experimental results, 
\begin{align} \begin{split}
\frac{1}{1080}\frac{(3J_1 -14J_1\rho_{1a})^2}{J_1-3J_1\rho_{1a}}
=\frac{e_+^2}{b_+}
\le
c_+.
\end{split} \end{align}

\subsection{A bound for $d_+$
}

\subsubsection{Without using information $c_+$}\label{d w/o c}
 
We obtain a bound for $d_+$ using information from $G_0(x_1)$ measurement without using information on $c_+$.

From the Eq. (\ref{J_1}),  
\begin{align} \begin{split}
\frac{1}{36}\frac{9+3\rho_{1a}-10\rho_{1b}}{3-3\rho_{1a}-2\rho_{1b}}
=\frac{8}{4a_++b_+}d_++\frac{a_+}{4a_++b_+}.
\end{split} \end{align}
If $d_+=0$, 
\begin{align} \begin{split}
0\le\frac{1}{36}\frac{9+3\rho_{1a}-10\rho_{1b}}{3-3\rho_{1a}-2\rho_{1b}}
\le\frac{1}{4}
\end{split} \end{align}
since $a_+,b_+\ge0$. 
So, if
\begin{align} \begin{split}
\frac{9+3\rho_{1a}-10\rho_{1b}}{3-3\rho_{1a}-2\rho_{1b}}
<0,
\end{split} \end{align}
then $d_+<0$; 
 if
\begin{align} \begin{split}
\frac{9+3\rho_{1a}-10\rho_{1b}}{3-3\rho_{1a}-2\rho_{1b}}
>9,
\end{split} \end{align}
then $d_+>0$;
%
%
%
 and if
\begin{align} \begin{split}
0\le\frac{9+3\rho_{1a}-10\rho_{1b}}{3-3\rho_{1a}-2\rho_{1b}}
\le 9,
\end{split} \end{align}
then, from Eq. (\ref{29}),
\begin{align} \begin{split}
-J_1\frac{3-3\rho_{1a}-2\rho_{1b}}{32}
\le d_+ \le
 J_1\frac{3-3\rho_{1a}-2\rho_{1b}}{32}.
\end{split} \end{align}
%
%
%
%

Furthermore, we can determine $d_+$ allowed region as follows.
Using the fact
\begin{align} \begin{split}
0\le\frac{a_+}{4a_++b_+}\le \frac{1}{4},
\end{split} \end{align}
which become the equation when $a_+=0$ and $b_+=0$, respectively,
\begin{align} \begin{split}
\frac{8}{4a_++b_+}d_+
\le
\frac{1}{36}\frac{9+3\rho_{1a}-10\rho_{1b}}{3-3\rho_{1a}-2\rho_{1b}}
\le
\frac{8}{4a_++b_+}d_++ \frac{1}{4}.
\end{split} \end{align}
The solution of this inequality about $d_+$ becomes 
\begin{align} \begin{split}
\frac{-9J_1+15J_1\rho_{1a}+4J_1\rho_{1b}
}{144}
\le
d_+
\le
\frac{9J_1+3J_1\rho_{1a}-10J_1\rho_{1b}}{288}.
\end{split} \end{align}
This becomes equality, when $b_+=0$ and $a_+=0$, respectively.

\subsubsection{ Using information $c_+$}

We obtain a bound for $d_+$ using information from $G_0(x_1)$ measurement with using information on $c_+$.

From Eq. (\ref{J_1}),
\begin{align} \begin{split}
a_+c_++8d_+c_+=\frac{J_1}{36}(9+3\rho_{1a}-10\rho_{1b})c_+.
\end{split} \end{align}
This becomes
\begin{align} \begin{split}
\frac{J_1}{36}(9+3\rho_{1a}-10\rho_{1b})c_+\ge d_+^2+8d_+c_+
\end{split} \end{align}
since $a_+c_+-d_+^2\ge0$ as explained in appendix \ref{6.1}.
Using $c_+>0$,
\begin{align} \begin{split}
\frac{J_1}{36c_+}(9+3\rho_{1a}-10\rho_{1b})
\ge \left(\frac{d_+}{c_+}+4\right)^2-16.
\end{split} \end{align}
The solution of this inequality about $d_+$ becomes
\begin{align} \begin{split}
-4c_+&-c_+\sqrt{\frac{J_1}{36c_+}(9+3\rho_{1a}-10\rho_{1b})+16}
\\&\le d_+ \le 
-4c_++c_+\sqrt{\frac{J_1}{36c_+}(9+3\rho_{1a}-10\rho_{1b})+16}.
\end{split} \end{align}
So we have another limit of $d_+$.


\subsection{A bound for $e_+$ 
}\label{b of e}

\subsubsection{Without using information $c_+$}\label{e w/o c}

We obtain a bound for $e_+$ using information from $G_0(x_1)$ measurement without using information on $c_+$.

Similar to the subsubsection \ref{d w/o c},
\begin{align} \begin{split}
\frac{4}{9}\frac{\rho_{1b}}{3-3\rho_{1a}-2\rho_{1b}}
=\frac{8}{4a_++b_+}e_++\frac{b_+}{4a_++b_+},
\end{split} \end{align}
and if $e_+=0$, 
\begin{align} \begin{split}
0\le\frac{4}{9}\frac{\rho_{1b}}{3-3\rho_{1a}-2\rho_{1b}}\le 1.
\end{split} \end{align}
This becomes equation when $b_+=0$ and $a_+=0$ respectively.
So,
if
\begin{align} \begin{split}
\frac{\rho_{1b}}{3-3\rho_{1a}-2\rho_{1b}}
<0,
\end{split} \end{align}
then $e_+<0$; 
 if
\begin{align} \begin{split}
\frac{\rho_{1b}}{3-3\rho_{1a}-2\rho_{1b}}
>\frac{9}{4},
\end{split} \end{align}
then $e_+>0$;
%
%
%
 and if
\begin{align} \begin{split}
0\le
\frac{\rho_{1b}}{3-3\rho_{1a}-2\rho_{1b}}
\le\frac{9}{4},
\end{split} \end{align}
then, from Eq. (\ref{29}),
\begin{align} \begin{split}
-J_1\frac{3-3\rho_{1a}-2\rho_{1b}}{8}
\le e_+ \le
 J_1\frac{3-3\rho_{1a}-2\rho_{1b}}{8}.
\end{split} \end{align}
%
%
%
%

Furthermore, we can determine $e_+$ allowed region as follows.
Using the fact
\begin{align} \begin{split}
0\le\frac{b_+}{4a_++b_+}\le 1,
\end{split} \end{align}
which become the equation when $b_+=0$ and $a_+=0$, respectively,
\begin{align} \begin{split}
\frac{8}{4a_++b_+}e_+
\le
\frac{4}{9}\frac{\rho_{1b}}{3-3\rho_{1a}-2\rho_{1b}}
\le
\frac{8}{4a_++b_+}e_++1.
\end{split} \end{align}
The solution of this inequality about $e_+$ becomes 
\begin{align} \begin{split}
\frac{J_1\rho_{1b}}{18}-\frac{3J_1-3J_1\rho_{1a}-2J_1\rho_{1b}}{8}
\le e\le
\frac{J_1\rho_{1b}}{18}.
\end{split} \end{align}
This becomes equality, when $a_+=0$ and $b_+=0$, respectively.

\subsubsection{ Using information $c_+$}

We obtain a bound for $e_+$ using information from $G_0(x_1)$ measurement with using information on $c_+$.

\label{b of e2}
Similar to previous subsection, from Eq. (\ref{J_1}),
\begin{align} \begin{split}
b_+c_++8e_+c_+=\frac{4}{9}J_1\rho_{1b}c_+.
\end{split} \end{align}
This becomes
\begin{align} \begin{split}
\frac{4}{9}J_1\rho_{1b_+}c_+\ge e_+^2+8e_+c_+
\end{split} \end{align}
since $b_+c_+-e_+^2\ge0$ as explained in appendix \ref{6.1}.
Using $c_+>0$,
\begin{align} \begin{split}
\frac{4}{9}\frac{J_1\rho_{1b}}{c_+}\ge \left(\frac{e_+}{c_+}+4\right)^2-16.
\end{split} \end{align}
The solution of this inequality about $e_+$ becomes
\begin{align} \begin{split}
-4c_+-2c_+\sqrt{\frac{J_1\rho_{1b}}{9c_+}+4}
\le e_+ \le -4c_++2c_+\sqrt{\frac{J_1\rho_{1b}}{9c_+}+4}.
\end{split} \end{align}
So we have another limit of $e_+$.

\section{$G_0(x_3)$: Energy Dependence of $\mu_3$ } \label{Energymu3}
   In the previous section, we have discussed how the values of parameters $c_+$, $J_1$, $\rho_{1a}$ and $\rho_{1b}$ can be obtained from  $G_0$ but we have not determined the values of $a_+$, $b_+$, $d_+$ and $e_+$.
    This will be the subject of this section.

\begin{figure}[htbp]
  \begin{center}
    \includegraphics[keepaspectratio=true,height=55mm]{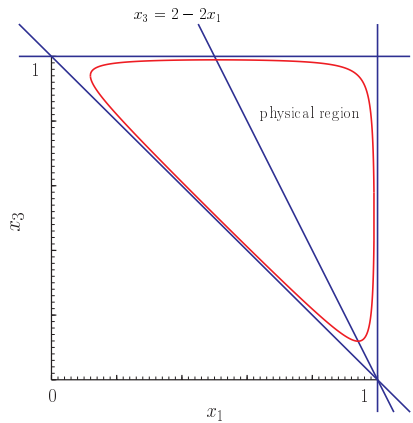}
  \end{center}
  \caption{The physical region for $\tau^+ \to \mu_1 \mu_2\mu_3$ is the area which is included in curbed line and besides the right side of the line $x_3=2-2x_1$.   
  Near the  line $x_1=1$, the differential branching ratio is singular since the intermediate state photon becomes real.}
\label{singular}
\end{figure}

    In the previous section, we gave $x_1$ dependence for $G_0$. In this section, to determine the values of $a_+$, $b_+$, $d_+$ and $e_+$, we give $x_3$ dependence of differential branching ratio.
    As in the previous section, we do not discuss the polarizations.
    For most of the analysis, muon mass dependence can be neglected.
    But, as we shall see, in some part of the phase space, $m_\mu$ dependence must be taken into account.

   To study the formula of the $G_0$ as function of $x_3$, we give the following prescription.
    We can get the $x_3$ dependence of $G_0(x_1,x_2)$ by using $x_1+x_2+x_3=2$. 
    From the condition $x_1\ge x_2$ and the relation $x_1+x_2+x_3=2$,
\begin{eqnarray}
x_1\ge 1-\frac{x_3}{2}.\label{斜線}
\end{eqnarray}
    So, if all muon masses can be neglected, we have  $1/2 \le  x_1 \le 1$ and $0\le  x_3\le 1$ as shown in Fig. \ref{singular}. 
   The integration of  $x_1$ in the $x_1\mbox{-} x_3$ plane is given by  
\begin{eqnarray} 
\in 1-\frac{x_3}{2} 1 dx_1 G_0(x_1,x_3). 
\end{eqnarray}

     When $x_1=1$ in massless limit of muons, the photon in Fig. \ref{fig:photo} becomes on shell and the propagator becomes singular. 
     This leads a divergence of $x_1=1$ in differential branching ratio for $\tau \to 3\mu$ decay.

    So we have to be more careful with the range of at $x_1$ including the muon mass dependence.
   The on shell constraint $p_2^2=m_\mu^2$ leads to   
\begin{eqnarray} 
\frac{\delta^2}{4}+1-x_1-x_3+\frac{x_1 x_3}{2}
-\frac{1}{2}\mathbf{\hat{p}}_1 \cdot \mathbf{\hat{p}}_3\sqrt{x_1^2-\delta^2}\sqrt{x_3^2-\delta^2} =0\label{relative},
\end{eqnarray}
    where $\delta \equiv 2m_\mu/m_\tau$.
    The true integration range for $x_1$ is given by  Eqs. (\ref{斜線}) and    (\ref{relative}) with $-1\le \mathbf{\hat{p}}_1 \cdot \mathbf{\hat{p}}_3 \le 1$ as shown in Fig.  \ref{singular}.
       We find it convenient to approximate the domain of integration  as $ 1-\frac{x_3}{2} \le  x_1 \le 1-(\frac{4\delta}{3})^2 $.
     Here, the upper bound of $x_1$ is decided  as follows.
    First, we calculate the total branching ratio of $ c_+$ sector integrated on the true physical region in $\mathcal{O}(\delta^2)$.
    Next, we calculate the total  branching ratio of $ c_+$ sector integrated on the approximated region.
    Finally, we set the approximated upper bound to match these two total branching ratios.
         This approximation allows us to avoid the divergent region.
    With this approximation, we give the  $x_3$ dependence of $G_0$ as
\begin{align} \begin{split}
G_0(x_3)= \in {1-\frac{x_3}{2}} {1-(\frac{4\delta}{3})^2} & dx_1 \ G_0(x_1,x_3)    
\\=&\ \frac{1}{2} J_3 x_3 \biggl\{ x_3(1-x_3)
               +\rho_{3a} x_3\bigl(x_3-\frac{2}{3}\bigr)
               +\rho_{3b} (x_3-1)\bigl(x_3-\frac{1}{3}\bigr) \biggr\}
               \\ 
&+4c_+ \left\{2x_3-3x_3^2+2(2x_3^2-2x_3+1)\log\bigl[\frac{ 3 \sqrt{x_3}}{4\delta }\bigr]\right\}, 
\end{split} \end{align}
where
\begin{align} \begin{split}\label{J_3}
J_3&=\frac{1}{3}(12a_+ +5b_+ +144d_+ +36e_+ ) 
             \\ J_3\rho_{3a}&=  b_+ +12e_+ 
            \\J_3 \rho_{3b}&= 48d_+ .  
\end{split} \end{align}

First, to determine $c_+$, it is convenient to define the function 
\begin{align} \begin{split}
F_3(x_3)= \frac{G_0(x_3)}{8\log \sqrt{x_3}}.
\end{split} \end{align}
By choosing the kinematics such that $x_3\to 0$ for the function $F_3(x_3)$, we can obtain $c_+$ as
\begin{align} \begin{split}
\left.F_3(x_3) \right|_{x_3\to 0}=c_+.
\end{split} \end{align}
Next, we define the function 
\begin{align} \begin{split}
F_4(x_3)& = 
 \frac{ G_0(x_3)  -(c_+,x_3 \ \mathrm{term})}{x_3}  
\\ &=  
 \frac{1}{2} J_3  \left\{ x_3(1-x_3)
               +\rho_{3a} x_3\bigl(x_3-\frac{2}{3}\bigr)
               +\rho_{3b} (x_3-1)\bigl(x_3-\frac{1}{3}\bigr) \right\},
\end{split} \end{align}
where 
\begin{eqnarray}
(c_+,x_3 \ \mathrm{term})=4c_+ \left\{2x_3-3x_3^2+2(2x_3^2-2x_3+1)\log\bigl [\frac{ 3 \sqrt{x_3}}{4\delta^2}\bigr]\right\}.
\end{eqnarray}
   From the function $F_4(x_3)$, we can obtain the values of $J_3$, $\rho_{3a}$ and $\rho_{3b}$ by fitting the experimental data as follows:

   \begin{align} \begin{split}
   \in 0 1 dx_3 F_4(x_3)&=\frac{1}{12}J_3 \\
 F_4(x_3)|_{x_3=1}&=\frac{1}{6}J_3\rho_{3a} \\
    F_4(x_3)|_{x_3=0}&=\frac{1}{6}J_3\rho_{3b}. 
    \end{split} \end{align}

   In addition to $J_1$, $\rho_{1a}$ and $ \rho_{1b}$ from previous section,  the parameters  $a_+$, $b_+$,  $d_+$ and $e_+$ can now be determined.
  For example, 
 \begin{align} \begin{split}
a_+&=\frac{1}{4} J_3\bigl(1+\frac{1}{3}\rho_{3a}-\rho_{3b}\bigr)-\frac{2}{9}J_1\rho_{1b}  \\
b_+&=\frac{4}{3} J_1\rho_{1b}-2 J_3\rho_{3a}  \\  
 d_+&= \frac{1}{48} J_3 \rho_{3b} \\
e_+&= \frac{1}{4} J_3\rho_{3a} -\frac{1}{9}  J_1\rho_{1b}. 
\end{split} \end{align}

From the differential cross section which are the functions of $x_1$ and $x_2$, $x_1$ and $x_3$ or $x_2$ and $x_3$, all we can determine are the quantities $a_+,b_+,c_+,d_+,e_+$.
Now, we will discuss some specific cases in which we can go further.

\section{Physics from $G_0(x_3)$ Distribution of $\mu_3$}\label{phys of 3}
Similar to the section \ref{phys of 1}, we have some bounds for $a_+$, $b_+$, $d_+$ and $e_+$ from the analysis of previous section.
\subsection{What can we conclude if one of $a_+$, $b_+$, $c_+$, $d_+$, $e_+$ $=0$}
If we impose that one of $a_+$, $b_+$, $ c_+ $, $ d_+ $ and $ e_+ $ is zero from other experimental results or supposing some specific models, the results are as follows.
\\ 
When $a_+=0$, also $d_+=0$ and 
\begin{align} \label{a=0,3}
 J_3&=\frac{1}{3}(5b_++36e_+)     \nonumber\\
  J_3\rho_{3a}&=b_++12e_+ \\
  J_3\rho_{3b}&=0.    \nonumber
 \end{align}
When $b_+=0$, also $e_+=0$ and  
\begin{align} \label{b=0,3}
 J_3&=4(a_++12d_+)     \nonumber\\
  J_3\rho_{3a}&=0 \\
  J_3\rho_{3b}&=48d_+.    \nonumber
 \end{align}
When $c_+=0$, also $d_+=e_+=0$ and
\begin{align}
 J_3&=\frac{1}{3}(12a_++5b_+)     \nonumber\\
  J_3\rho_{3a}&=b_+  \\
  J_3\rho_{3b}&=0.    \nonumber
\end{align}
When $d_+=0$,
\begin{align} \begin{split}
J_3&=\frac{1}{3}(12a_+ +5b_+  +36e_+ ) 
             \\ J_3\rho_{3a}&=  b_+ +12e_+ 
            \\J_3 \rho_{3b}&=0.  
\end{split} \end{align}
When $e_+=0$,
\begin{align} \begin{split}
J_3&=\frac{1}{3}(12a_+ +5b_+ +144d_+  ) 
             \\ J_3\rho_{3a}&=  b_+  
            \\J_3 \rho_{3b}&= 48d_+ .  
\end{split} \end{align}
If any one of $a_+$, $b_+$, $c_+$ or $e_+$ is zero, we can determine $a_+$, $b_+$, $c_+$, $d_+$ and $e_+$ from only the $x_3$ distribution.
However, if $d_+=0$, then we cannnot determine $a_+$, $b_+$ and $e_+$, independently. 

\subsection{Results if either $a_+\not=0$, $b_+\not=0$, $d_+\not=0$ or $e_+\not=0$
 }
Similarly,
When only $a_+\not=0$, 
\begin{align}
  J_3&=4a_+ \nonumber\\
\rho_{3a}&=0\\
 \rho_{3b}&=0.\nonumber
\end{align}
When only $b_+\not=0$, 
\begin{align}
  J_3&=\frac{5}{3}b_+ \nonumber\\
\rho_{3a}&=\frac{3}{5}\\
 \rho_{3b}&=0.\nonumber
\end{align}
When only $a_+,\ c_+,\ d_+\not=0$ but $a_+$ is negligible, 
\begin{align} \begin{split}
J_3&=48d_+ 
\\ \rho_{3a}&= 0 
\\ \rho_{3b}&= 1 .  
\end{split} \end{align}
When only $b_+,\ c_+,\ e_+\not=0$ but $b_+$ is negligible, 
\begin{align} \begin{split}
J_3&=12e_+ 
\\ J_3\rho_{3a}&= 1 
\\J_3 \rho_{3b}&= 0.  
\end{split} \end{align}
In these case, $\rho_{3a}$ and $\rho_{3b}$ are independent from $a_+$, $b_+$, $d_+$ and $e_+$.

The results in subsection \ref{6.8} and here is summed in the table below.

\begin{table}[htb]
 \caption{values of $J's$ and $\rho$'s in some cases}
 \begin{center}
  \begin{tabular}{|c||c|c|c||c|c|c|}
    \hline
     &$J_1$ & $\rho_{1a}$   & $\rho_{1b}$  &$J_3$ & $\rho_{3a}$   &  $\rho_{3b}$  \\
    \hline 
    \hline
  \raisebox{-1ex}{  (A)} & \raisebox{-1ex}{ $\frac{10}{3}a_+$} & \raisebox{-1ex}{ $\frac{3}{5}$}& \raisebox{-1ex}{ $0$}& \raisebox{-1ex}{ $4a_+$} & \raisebox{-1ex}{  $0$ }  & \raisebox{-1ex}{  $0$}  \\[2ex]
    \hline
    \raisebox{-1ex}{  (B)}& \raisebox{-1ex}{ $\frac{7}{3}b_+$} & \raisebox{-1ex}{ $\frac{3}{14}$}& \raisebox{-1ex}{ $\frac{27}{28}$}& \raisebox{-1ex}{ $\frac{5}{3}b_+$ }&  \raisebox{-1ex}{ $\frac{3}{5}$ }  & \raisebox{-1ex}{  $0$ }     \\[2ex]

        \hline
     \raisebox{-0ex}{     (C)}&$0$ &$-$&$-$&$0$ & $-$   & $-$      \\
        \hline
 \raisebox{-0ex}{  (D)}&$24d_+$ &$1$&$0$&$48d_+$ &$0$  & $1$      \\
        \hline 
  \raisebox{-1ex}{ (E)}& \raisebox{-1ex}{ $18e_+$} & \raisebox{-1ex}{ $\frac{1}{3}$}& \raisebox{-1ex}{ $1$}& \raisebox{-1ex}{ $12e_+$} & \raisebox{-1ex}{ $1$ }  &  \raisebox{-1ex}{ $0$}      \\[2ex]
        \hline
 \end{tabular}
  \\
\begin{flushleft}
\hspace{4em} (A): all but $g_1$, $g_2$, $g_3$, $g_4$ are vanishing\\
\hspace{4em}   (B): all but $g_5$, $g_6$ are vanishing\\ 
\hspace{4em}   (C): all but $eA_R$, $eA_L$ are vanishing \\
\hspace{4em}   (D): all but $Re[g_3eA_L^*]+Re[g_4eA_R^*]$, $eA_R$, $eA_L$ are vanishing \\
\hspace{4em}  (E): all but $Re[g_5eA_L^*]+Re[g_6eA_R^*]$, $eA_R$, $eA_L$ are vanishing 
 \end{flushleft}
  \end{center}
\end{table}

\subsection{What happens if $\tau\to\mu\gamma$ dominates  }

Similar to subsection \ref{c>>,1}, in this case,
\begin{align} \begin{split}
J_3&=12(4d_++e_+)\\
J_3\rho_{3a}&=12e_+\\
J_3\rho_{3b}&=48d_+.
\end{split} \end{align}
From Eqs. (\ref{Eq h}) and (\ref{Eq h2}) in appendix \ref{6.1}, 
\begin{align} \begin{split}
a_+&\ge\frac{d_+^2}{c_+}=\frac{1}{c_+}\left(\frac{J_3\rho_{3b}}{48}\right)^2\\
b_+&\ge\frac{e_+^2}{c_+}=\frac{1}{c_+}\left(\frac{J_3\rho_{3a}}{12}\right)^2.
\end{split} \end{align}
So, we can determine  $a_+$ and $b_+$ lower limit, though we can't determine  $a_+$ and $b_+$, directly.

We note here that since the relation $J_3\rho_{3b}=48d_+$ is always true, $a_+$ lower limit is always determined.

\subsection{Lower limit of $c_+$ when $\tau\to\mu\gamma$ is highly suppressed}

We give three types of $c_+$ lower limits. 
Especially, these are very useful when $\tau\to\mu\gamma$ is highly suppressed compared with $\tau\to 3\mu$ and $c_+$ cannot determine directly.

First, we give the relations,
\begin{align} \begin{split}\label{J_32}
3J_3-5J_3\rho_{3a}-3J_3\rho_{3b}&=12(a_+-2e_+)\\
3J_3-3J_3\rho_{3a}-3J_3\rho_{3b}&=2(6a_++b_+)\\
J_3\rho_{3a}&=b_++12e_+\\
J_3\rho_{3b}&=48d_+
\end{split} \end{align}
from Eqs. (\ref{J_3}).

Using Eqs. (\ref{J_32}) and (\ref{Eq h}) and the fact that $a_+,\ b_+ \ge 0$,
\begin{align} \begin{split}
\frac{(J_3\rho_{3b})^2}{576(J_3-J_3\rho_{3a}-J_3\rho_{3b})}=\frac{6d_+^2}{6a_++b_+}\le\frac{d_+^2}{a_+}\le c_+.
\end{split} \end{align}

Using Eqs. (\ref{J_32}),
\begin{align} \begin{split}
\frac{(J_3\rho_{3b})^2}{192(3J_3-5J_3\rho_{3a}-3J_3\rho_{3b})}=\frac{d_+^2}{a_+-2e_+}.
\end{split} \end{align}
 Here, if $c_+\ll b_+$, then $J_3\rho_{3a}=b_++12e_+=\mathcal{O}( b_+)$ and $b_+\ge \mathcal{O}( |12e_+|)$.
 So, if $(3J_3-5J_3\rho_{3a}-3J_3\rho_{3b})/12 \gse J_3\rho_{3a}   $, this means $ a_+-2e_+ \ge\mathcal{O}( b_+) $ and then $a_+-2e_+\simeq a_+$. 
 So, using the relation (\ref{Eq h}),
\begin{align} \begin{split}
\frac{(J_3\rho_{3b})^2}{192(3J_3-5J_3\rho_{3a}-3J_3\rho_{3b})}\lse c_+.
\end{split} \end{align}

If 
\begin{align} \begin{split}
3J_3-5J_3\rho_{3a}-3J_3\rho_{3b}=12(a_+-2e_+)<0,
\end{split} \end{align}
$b_+\gg c_+$ and  there is no special cancelation between $a_+$ and $2e_+$, then $a_+-2e_+=\mathcal{O}(-2e_+)$ and $a_+\le \mathcal{O}(2e_+)\ll b_+/6$.
So,
\begin{align} \begin{split}\label{denom c}
\frac{(3J_3-5J_3\rho_{3a}-3J_3\rho_{3b})^2}{864(J_3-J_3\rho_{3a}-J_3\rho_{3b})}
=\frac{(a_+-2e_+)^2}{4(6a_++b_+)}
=\mathcal{O}\left(\frac{e_+^2}{b_+}\right)
\le \mathcal{O}(c_+).
\end{split} \end{align}
We note here that $6a_++b_+$ is similar to $b_++12e_+$ in this situation. 
So, we have similar result if the denominator of left hand side in Eq. (\ref{denom c}),     $
J_3-J_3\rho_{3a}-J_3\rho_{3b}=2(6a_++b_+)/3
 $ is exchanged by $ 
2J_3\rho_{3a}/3=2(b_++12e_+)/3.$

From (\ref{b=0,3}) and (\ref{Eq h}), if we suppose that $b_+=0$ from some special models or other experimental results, 
\begin{align} \begin{split}
\frac{1}{24^2}\frac{(J_3\rho_{3b})^2}{J_3-J_3\rho_{b}}
=\frac{d_+^2}{a_+}
\le
c_+.
\end{split} \end{align}
Similarly, from (\ref{a=0,3}) and (\ref{Eq h2}), if we suppose that $a_+=0$ from some special models or other experimental results, 
\begin{align} \begin{split}
\frac{1}{864}\frac{(3J_3 -5J_3\rho_{1a})^2}{J_3-J_3\rho_{1a}}
=\frac{e_+^2}{b_+}
\le
c_+.
\end{split} \end{align}

\subsection{A bound for $e_+$ }

\subsubsection{Without using information $c_+$}
We obtain a bound for $e_+$ using information from $G_0(x_1)$ measurement without using information on $c_+$.

Similarly to the subsection \ref{b of e},
\begin{align} \begin{split}
\frac{2\rho_{3a}}{3-3\rho_{3a}-3\rho_{3b}}
=\frac{12}{6a_++b_+}e_++\frac{b_+}{6a_++b_+}, 
\end{split} \end{align}
and if $e_+=0$, 
\begin{align} \begin{split}
0\le\frac{2\rho_{3a}}{3-3\rho_{3a}-3\rho_{3b}}\le1.
\end{split} \end{align}
This becomes equation when $b_+=0$ and $a_+=0$ respectively.
So,
if
\begin{align} \begin{split}
\frac{\rho_{3a}}{1-\rho_{3a}-\rho_{3b}}
<0,
\end{split} \end{align}
then $e_+<0$; 
 if
\begin{align} \begin{split}
\frac{\rho_{3a}}{1-\rho_{3a}-\rho_{3b}}>\frac{3}{2},
\end{split} \end{align}
then $e_+>0$;
%
%
%
 and if
\begin{align} \begin{split}
0\le
\frac{\rho_{3a}}{1-\rho_{3a}-\rho_{3b}}
\le
\frac{3}{2},
\end{split} \end{align}
then, from Eqs. (\ref{J_32}), 
\begin{align} \begin{split}
-J_3\frac{1-\rho_{3a}-\rho_{3b}}{8}
\le e_+ \le
 J_3\frac{1-\rho_{3a}-\rho_{3b}}{8}.
\end{split} \end{align}
%
%
%
%

Furthermore, we can determine $e_+$ allowed region as follows.
Using the fact
\begin{align} \begin{split}
0\le\frac{b_+}{6a_++b_+}\le1,
\end{split} \end{align}
which become the equation when $b_+=0$ and $a_+=0$, respectively,
\begin{align} \begin{split}
\frac{12}{6a_++b_+}e_+
\le\frac{2\rho_{3a}}{3-3\rho_{3a}-3\rho_{3b}}
\le\frac{12}{6a_++b_+}e_++1. 
\end{split} \end{align}
The solution of this inequality about $e_+$ becomes 
\begin{align} \begin{split}
\frac{-3J_3+5J_3\rho_{3a}+3J_3\rho_{3b}}{24}
\le e_+ \le
\frac{J_3\rho_{3a}}{12}. 
\end{split} \end{align}
This becomes equality, when $a_+=0$ and $b_+=0$, respectively.

\subsubsection{Using information $c_+$}
We obtain a bound for $e_+$ using information from $G_0(x_1)$ measurement with using information on $c_+$.

Similar to subsection \ref{b of e2}, from Eq. (\ref{J_3}),
\begin{align} \begin{split}
b_+c_++12e_+c_+=J_3\rho_{3b}c_+.
\end{split} \end{align}
This becomes
\begin{align} \begin{split}
J_3\rho_{3b_+}c_+\ge e_+^2+12e_+c_+
\end{split} \end{align}
since $b_+c_+-e_+^2\ge0$ as explained in appendix \ref{6.1}.
Using $c_+>0$,
\begin{align} \begin{split}
\frac{J_3\rho_{3b}}{c_+}\ge \left(\frac{e_+}{c_+}+6\right)^2-36.
\end{split} \end{align}
The solution of this inequality about $e_+$ becomes
\begin{align} \begin{split}
-6c_+-c_+\sqrt{\frac{J_3\rho_{3b}}{c_+}+36}
\le e_+ \le -6c_++c_+\sqrt{\frac{J_3\rho_{3b}}{c_+}+36}
\end{split} \end{align}
This becomes equality when $b_+c_+-e_+^2=0$.
So we have another limit of $e_+$.

\section{Physics Implications from $G_0(x_1)$ and $G_0(x_3)$ Distributions}\label{E 1,3}

From the analysis of both of $G_0(x_1)$ and $G_0(x_3)$ distributions,
we can determine not only $a_+$, $b_+$, $c_+$, $d_+$ and $e_+$ but also  in more detail in some suitable cases as explained in this section.

Introducing new real parameters $\{r_1,r_2,r_3,r_4,r_5,r_6,r_R,r_L\}\ge0$ and
 $  2\pi > \{\theta_1,\theta_2,$ $\theta_3, \theta_4, \theta_5, \theta_6, \theta_R, \theta_L\}\ge 0 $, the effective coupling constants can be explained as  
\begin{align} 
\label{rtheta}
g_1&=r_1 e^{i\theta_1}& \hspace{-3em}
g_2&=r_2 e^{i\theta_2}  \nonumber\\
g_3&=r_3 e^{i\theta_3}&\hspace{-3em}
g_4&=r_4 e^{i\theta_4}\nonumber\\
g_5&=r_5 e^{i\theta_5}&\hspace{-3em}
g_6&=r_6 e^{i\theta_6} \\
eA_R&=r_R e^{i\theta_R}&\hspace{-3em}
eA_L&=r_L e^{i\theta_L}.\nonumber
\end{align}
Then,
\begin{align} \begin{split}
a_+&=\frac{1}{16}(r_1^2+r_2^2)+r_3^2+r_4^2  \\
b_+&=r_5^2+r_6^2  \\
 c_+& =r_R^2+r_L^2 \\
 d_+&=-r_3 r_L\cos(\theta_3-\theta_L)- r_4 r_R\cos(\theta_4-\theta_R)\\
 e_+&=- r_6 r_R\cos(\theta_6-\theta_R)-r_5 r_L\cos(\theta_5-\theta_L).
\end{split} \end{align}

\subsection{What can we say if $a_+c_+-d_+^2=0$}\label{6.2}

From Eq. (\ref{Eq h}) in appendix \ref{6.1},
$a_+c_+-d_+^2=0$ only if 
\begin{numcases}
{}
\ g_3e A_R=g_4 eA_L    \label{3R=4L}
\\ \ Im[g_3 eA_L^*]+Im[g_4 eA_R^*]=0   \label{Im+Im=0}
\\ \  g_1=g_2=0. \label{g1=g2=0}
\end{numcases}
In that case,
\begin{align} \begin{split}
r_3 r_Re^{i(\theta_3+\theta_R)}=r_4 r_L e^{i(\theta_4+\theta_L)}
\end{split} \end{align}
from the relation (\ref{3R=4L}).
This means 
 
\begin{eqnarray}
\left\{\ 
  \begin{array}{l}
r_3 r_R=r_4 r_L       \\
   \theta_3-\theta_L=\theta_4-\theta_R,\ \theta_4-\theta_R \pm2\pi.    \\
  \end{array}
\right.
\end{eqnarray}

Then, from the relation (\ref{Im+Im=0}),
\begin{align} \begin{split}
&r_3 r_L\sin(\theta_3-\theta_L)
 +r_4 r_R\sin(\theta_4-\theta_R)
 \\&=r_3 r_L\sin(\theta_3-\theta_L)
 +r_4 r_R\sin(\theta_3-\theta_L)
\\&=(r_3 r_L+r_4 r_R)\sin(\theta_3-\theta_L)=0.
\end{split} \end{align}
This means
\begin{align} \begin{split}
r_3 r_L&+r_4 r_R=0
\\&\mathrm{and/or}
\\\theta_3&-\theta_L=0, \ \pm\pi.
\end{split} \end{align}
Here, $r_3 r_L+r_4 r_R\not=0$ because of the conditions (\ref{3}) and (\ref{g1=g2=0}).
Then $\theta_3-\theta_L=\theta_4-\theta_R=0,\ \pm\pi$ and
\begin{align} \begin{split}
 d_+=\mp(r_3 r_L+ r_4 r_R).
\end{split} \end{align}
The sign of right hand side is minus if $\theta_3-\theta_L=\theta_4-\theta_R=0$ and plus if $\theta_3-\theta_L=\theta_4-\theta_R=\pm\pi$.

In these cases,  
 \begin{eqnarray}\label{Eq f}
 \left\{\ 
  \begin{array}{l}
  a_+=r_3^2+r_4^2\\
 c_+=r_R^2+r_L^2\\
 d_+=\mp(r_3 r_L+ r_4 r_R)\\
 r_3r_R=r_4r_L.
     \\
\end{array}
\right.
\end{eqnarray}
One of these conditions is dependent on others.
For example, one condition $c_+=r_R^2+r_L^2$ can be expressed using other conditions as 
 \begin{eqnarray}
 \left\{\ 
  \begin{array}{l}
  \frac{d_+^2}{a_+}=\frac{(r_3 r_L+ r_4 r_R)^2}{r_3^2+r_4^2}
=\frac{(\frac{r_4r_L}{r_R} r_L+ r_4 r_R)^2}{\frac{r_4^2r_L^2}{r_R^2}+r_4^2}
=r_R^2+r_L^2\\
\frac{d_+^2}{a_+}=c_+\\
\end{array}
\right.
\end{eqnarray}
 since $a_+c_+-d_+^2=0$.
So, there are only three independent conditions in (\ref{Eq f}).
Using one of $r_3$, $r_4$, $r_R$ and $r_L$, we can express others. 
For instance, if we know $r_R$ from other experiments or some special models, we can represent other coupling constants as
\begin{align} \begin{split}\label{renritu}
 r_L&=\sqrt{c_+-r_R^2}\\
 r_4&=\frac{|d_+|r_R}{c_+}\\
 r_3&=\sqrt{a_+}\sqrt{1-\frac{r_R^2}{c_+}}.
\end{split} \end{align}
Here, we note that if $\{r_3,r_4,r_R,r_L\}\not=0$,  
\begin{align} \begin{split}
\frac{a_+}{c_+}&=\frac{|g_3|^2}{|eA_L|^2}=\frac{|g_4|^2}{|eA_R|^2}\\
\frac{|d_+|}{c_+}&=\frac{|g_3|}{|eA_L|}=\frac{|g_4|}{|eA_R|}
\end{split} \end{align}
from the relation (\ref{3R=4L}).

If $r_3=0$, then $r_L=0$ from the relations (\ref{3}), (\ref{3R=4L}) and (\ref{g1=g2=0}).
Similarly, $r_R=0$ if $r_4=0$; $r_3=0$ if $r_L=0$; and $r_4=0$ if $r_R=0$.

\subsection{What if it so happens that $a_+c_+-d_+^2>0$}\label{6.3}

When $a_+c_+-d_+^2>0$, at least, one of the relations
\begin{align} \begin{split}
g_3 A_R-g_4 A_L&\not=0    
\\ Im[g_3 A_L^*]+Im[g_4 A_R^*]&\not=0  
\\  g_1&\not=0
\\  g_2&\not=0
\end{split} \end{align}
is satisfied.

 First, if $g_3 A_R-g_4 A_L \not=0$, then $g_3 \not=g_4 $ and/or $ A_R \not= A_L$.   This leads parity violation which is defined in Appendix \ref{CPT}.     
 Next, if $Im[g_3 A_L^*]+Im[g_4 A_R^*] \not=0$, then the relative phases of one of $g_3 A_L^*$ and/or $g_4 A_R^*$ have nonzero values.
This leads CP violation.
 If $g_1\not=0$ or $g_2 \not=0$, the scalar current and/or pseudo scalar current exists.

 \subsection{What can we say if $b_+c_+-e_+^2=0$}\label{7.2}

From Eq. (\ref{Eq h2}) in appendix \ref{6.1},
$b_+c_+-e_+^2=0$ only if 
\begin{numcases}
{}
\ g_5e A_R=g_6 eA_L    \label{5R=6L}
\\ \ Im[g_5 eA_L^*]+Im[g_6 eA_R^*]=0   \label{Im+Im=02}
\end{numcases}
In that case,
\begin{align} \begin{split}
r_5 r_Re^{i(\theta_5+\theta_R)}=r_6 r_L e^{i(\theta_6+\theta_L)}
\end{split} \end{align}
from the relation (\ref{5R=6L}).
This means 
 
\begin{eqnarray}
\left\{\ 
  \begin{array}{l}
r_5 r_R=r_6 r_L       \\
   \theta_5-\theta_L=\theta_6-\theta_R,\ \theta_5-\theta_R \pm2\pi.    \\
  \end{array}
\right.
\end{eqnarray}

Then, from the relation (\ref{Im+Im=02}),
\begin{align} \begin{split}
&r_5 r_L\sin(\theta_5-\theta_L)
 +r_6 r_R\sin(\theta_6-\theta_R)
 \\&=r_5 r_L\sin(\theta_5-\theta_L)
 +r_6 r_R\sin(\theta_5-\theta_L)
\\&=(r_5 r_L+r_6 r_R)\sin(\theta_5-\theta_L)=0.
\end{split} \end{align}
This means
\begin{align} \begin{split}
r_5 r_L&+r_6 r_R=0
\\&\mathrm{and/or}
\\\theta_5&-\theta_L=0, \ \pm\pi.
\end{split} \end{align}
Here, $r_5 r_L+r_6 r_R\not=0$ because of the conditions (\ref{32}).
Then $\theta_5-\theta_L=\theta_6-\theta_R=0,\ \pm\pi$ and
\begin{align} \begin{split}
 e_+=\mp(r_5 r_L+ r_6 r_R).
\end{split} \end{align}
The sign of right hand side is minus if $\theta_5-\theta_L=\theta_6-\theta_R=0$ and plus if $\theta_5-\theta_L=\theta_6-\theta_R=\pm\pi$.

In these cases,  
 \begin{eqnarray}\label{Eq f2}
 \left\{\ 
  \begin{array}{l}
  b_+=r_5^2+r_6^2\\
 c_+=r_R^2+r_L^2\\
 e_+=\mp(r_5 r_L+ r_6 r_R)\\
 r_5r_R=r_6r_L.
     \\
\end{array}
\right.
\end{eqnarray}
One of these conditions is dependent on others.
For example, one condition $c_+=r_R^2+r_L^2$ can be expressed using other conditions as 
 \begin{eqnarray}
 \left\{\ 
  \begin{array}{l}
  \frac{e_+^2}{b_+}=\frac{(r_5 r_L+ r_6 r_R)^2}{r_5^2+r_6^2}
=\frac{(\frac{r_6r_L}{r_R} r_L+ r_6 r_R)^2}{\frac{r_6^2r_L^2}{r_R^2}+r_6^2}
=r_R^2+r_L^2\\
\frac{e_+^2}{b_+}=c_+\\
\end{array}
\right.
\end{eqnarray}
 since $b_+c_+-e_+^2=0$.
So, there are only three independent conditions in (\ref{Eq f2}).
Using one of $r_5$, $r_6$, $r_R$ and $r_L$, we can express others. 
For instance, if we know $r_R$ from other experiments or some special models, we can represent other coupling constants as
\begin{align} \begin{split}\label{renritu2}
 r_L&=\sqrt{c_+-r_R^2}\\
 r_6&=\frac{|e_+|r_R}{c_+}\\
 r_5&=\sqrt{b_+}\sqrt{1-\frac{r_R^2}{c_+}}.
\end{split} \end{align}
Here, we note that if $\{r_5,r_6,r_R,r_L\}\not=0$,  
\begin{align} \begin{split}
\frac{b_+}{c_+}&=\frac{|g_5|^2}{|eA_L|^2}=\frac{|g_6|^2}{|eA_R|^2}\\
\frac{|e_+|}{c_+}&=\frac{|g_5|}{|eA_L|}=\frac{|g_6|}{|eA_R|}
\end{split} \end{align}
from the relation (\ref{5R=6L}).

If $r_5=0$, then $r_L=0$ from the relations (\ref{32}) and (\ref{5R=6L}).
Similarly, $r_R=0$ if $r_6=0$; $r_5=0$ if $r_L=0$; and $r_6=0$ if $r_R=0$.

\subsection{What if it so happens that $b_+c_+-e_+^2>0$}\label{7.3}

When $b_+c_+-e_+^2>0$, at least, one of the relations
\begin{align} \begin{split}
g_5 A_R-g_6 A_L&\not=0    
\\ Im[g_5 A_L^*]+Im[g_6 A_R^*]&\not=0  
\end{split} \end{align}
is satisfied.

 First, if $g_5 A_R-g_6 A_L \not=0$, then $g_5 \not=g_6 $ and/or $ A_R \not= A_L$.   This leads parity violation which is defined in Appendix \ref{CPT}.     
 Next, if $Im[g_5 A_L^*]+Im[g_6 A_R^*] \not=0$, then the relative phases of one of $g_5 A_L^*$ and/or $g_6 A_R^*$ have nonzero values.
This leads CP violation. 

\subsection{What if one of $A_L,A_R,g_3,g_4g_5,g_6=0$}\label{6.4}

We reveal here that even if $a_+c_+-d_+^2>0$ or $b_+c_+-e_+^2>0$, we can determine the lower limit of $|g_3|$, $|g_4|$, $|g_5|$, $|g_6|$, $|eA_R|$ and $|eA_L|$ in the case $A_L=0$, $A_R=0$, $A_L=0$, $A_R=0$, $g_3$ or $g_5=0$ and $g_4$ or $g_6=0$, respectively. 
If we restrict that $A_L=0$ from other observation for example $\tau\to\mu\gamma$ decay or some specific models,
\begin{align} \begin{split}\label{80}
\frac{d_+^2}{c_+}&=r_4^2\cos^2(\theta_4-\theta_R)\le |g_4|^2,\\
\frac{e_+^2}{c_+}&=r_6^2\cos^2(\theta_6-\theta_R)\le |g_6|^2.
\end{split} \end{align}
So, the lower limit of $|g_4|$, $|g_6|$ is determined.
Similarly, if $A_R=0$, 
\begin{align} \begin{split}
\frac{d_+^2}{c_+}&=r_3^2\cos^2(\theta_3-\theta_L)\le |g_3|^2,\\
\frac{e_+^2}{c_+}&=r_5^2\cos^2(\theta_5-\theta_L)\le |g_5|^2;
\end{split} \end{align}
if $g_3=0$,
\begin{align} \begin{split}
\frac{d_+^2}{a_+}
=\frac{r_4^2}{r_4^2+\frac{r_1^2+r_2^2}{16}}r_R^2\cos^2(\theta_4-\theta_R)
\le r_R^2\cos^2(\theta_4-\theta_R)
\le |eA_R|^2;
\end{split} \end{align}
if $g_5=0$,
\begin{align} \begin{split}
\frac{e_+^2}{b_+}
=
r_R^2\cos^2(\theta_6-\theta_R)
\le |eA_R|^2;
\end{split} \end{align}
and if $g_4=0$,
\begin{align} \begin{split}
\frac{d_+^2}{a_+}
=\frac{r_3^2}{r_3^2+\frac{r_1^2+r_2^2}{16}}r_L^2\cos^2(\theta_3-\theta_L)
\le r_L^2\cos^2(\theta_3-\theta_L)
\le |eA_L|^2;
\end{split} \end{align}
if $g_6=0$,
\begin{align} \begin{split}
\frac{e_+^2}{b_+}
=
 r_L^2\cos^2(\theta_5-\theta_L)
\le |eA_L|^2.
\end{split} \end{align}
The lower limits of $|g_4|$, $|g_6|$, $|g_3|$, $|g_5|$, $|eA_R|$ and $|eA_L|$ are determined in each case.

\subsection{What can we say about parity and CP symmetries}\label{6.5}
If parity or charge symmetry exists,
\begin{align} \begin{split}
r_1&=r_2\\
r_3&=r_4\\
r_5&=r_6\\
\theta_3&=\pm\theta_4\\
\theta_5&=\pm\theta_6\\
r_R&=r_L\\
\theta_R&=\pm\theta_L.
\end{split} \end{align}
So, 
\begin{align} \begin{split}\label{6.5-1}
a_+ c_+ - d_+^2=\frac{r_1^2r_L^2}{4}+4r_3^2r_L^2\sin^2(\theta_3-\theta_L)
\end{split} \end{align}
from Eq. (\ref{Eq h}), and 
\begin{align} \begin{split}\label{6.5-12}
b_+ c_+ - e_+^2=4r_5^2r_L^2\sin^2(\theta_5-\theta_L)
\end{split} \end{align}
from Eq. (\ref{Eq h2}).
If $a_+c_+-d_+^2\not=0$, then $|g_1|=|g_2|\not=0$ and/or CP is violated.
If $b_+c_+-e_+^2\not=0$, then CP is violated.

If both of parity and CP symmetries exist,
\begin{align} \begin{split}
r_1&=r_2\\
r_3&=r_4\\
r_5&=r_6\\
\theta_3&=\theta_4=0, \pi\\
\theta_5&=\theta_6=0, \pi\\
r_R&=r_L\\
\theta_R&=\theta_L=0, \pi.
\end{split} \end{align}
So, 
\begin{align} \begin{split}
a_+ c_+ - d_+^2=\frac{r_1^2r_L^2}{4}
\end{split} \end{align}
from Eq. (\ref{Eq h}), and
\begin{align} \begin{split}
b_+ c_+ - e_+^2=0
\end{split} \end{align}
from Eq. (\ref{Eq h2}).
When $a_+c_+-d_+^2\not=0$, $|g_1|=|g_2|\not=0$.

\subsection{What happens if $ Br(\tau\to\mu\gamma)$ times the fine structure constant is much smaller than $Br(\tau\to 3\mu)$}
In the case $c_+\ll \mathrm{others}$, we can't determine $c_+$ directly.
However, using the relations (\ref{Eq h}) and (\ref{Eq h2}), 
\begin{align} \begin{split}
c_+&\ge \frac{d_+^2}{a_+},\\
c_+&\ge \frac{e_+^2}{b_+}.
\end{split} \end{align}
So, we can determine $c_+$ lower limit.

This method is very useful to predict $\tau \to \mu \gamma$ branching ratio in the case that $\tau \to 3 \mu$ is detected while $\tau \to \mu \gamma$ has not been detected, yet.

 \section{Angular Distribution, $G_i^s(x_1)$ }\label{sec 5}

\begin{figure}[htbp]
  \begin{center}
    \includegraphics[keepaspectratio=true,height=50mm]{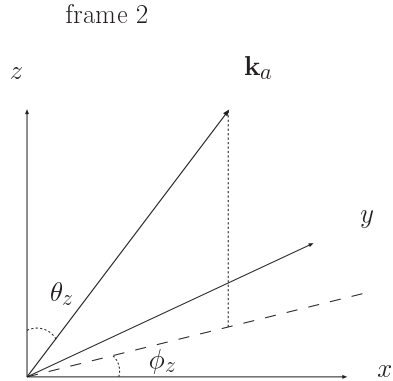}
  \end{center}
  \caption{The polar coordinate of frame 2. The North pole is $z$ direction.}
  \label{fig:frame2thetaphi-z.eps}
\end{figure}

\begin{figure}[htbp]
 \begin{minipage}[t]{.47\textwidth}
     \includegraphics[keepaspectratio=true,height=50mm]{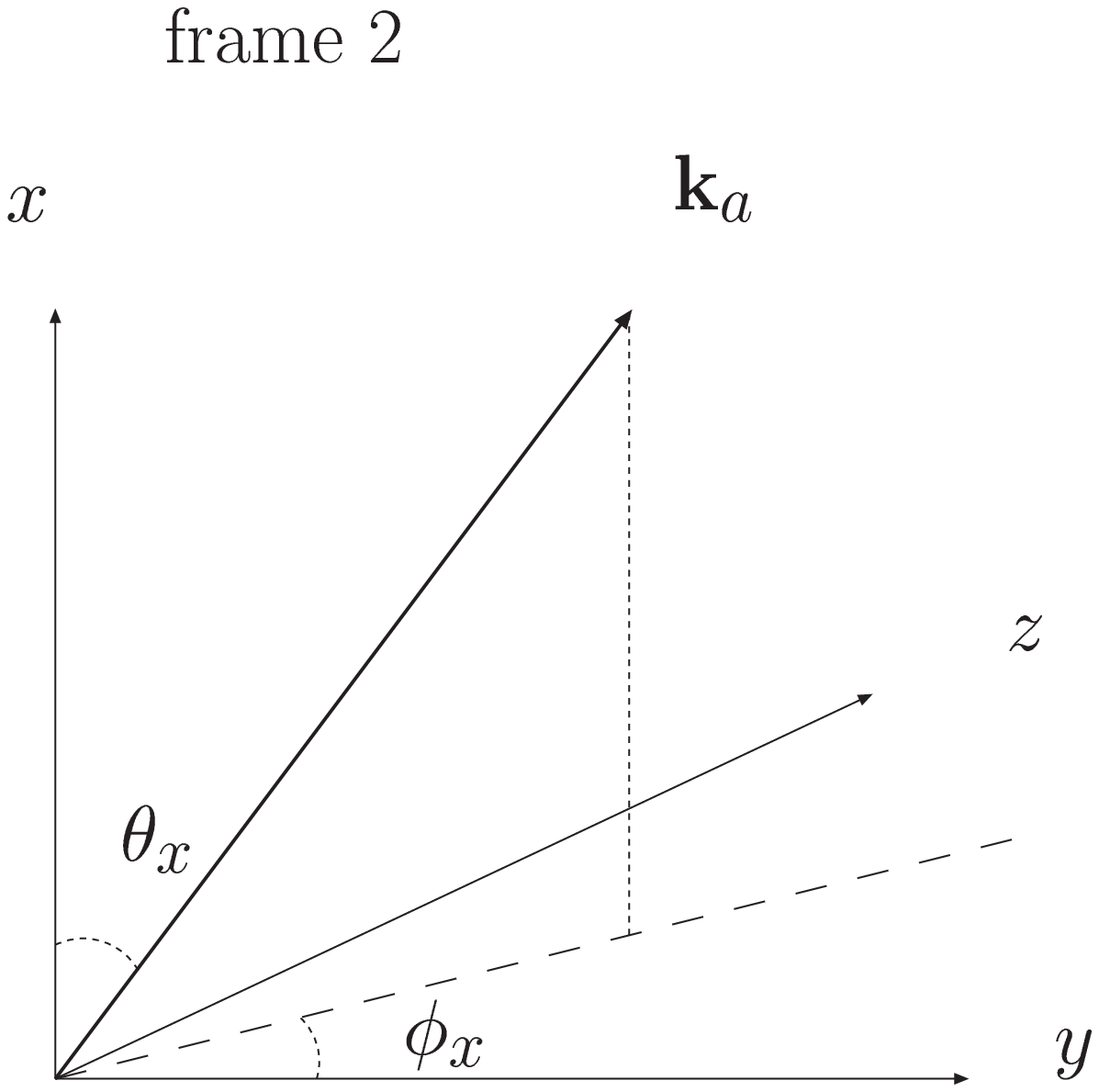}
  \caption{The polar coordinate of frame 2. The North pole is $x$ direction.}
  \label{fig:frame2thetaphi-x.eps}
   \end{minipage}
\hfill
 \begin{minipage}[t]{.47\textwidth}
    \includegraphics[keepaspectratio=true,height=50mm]{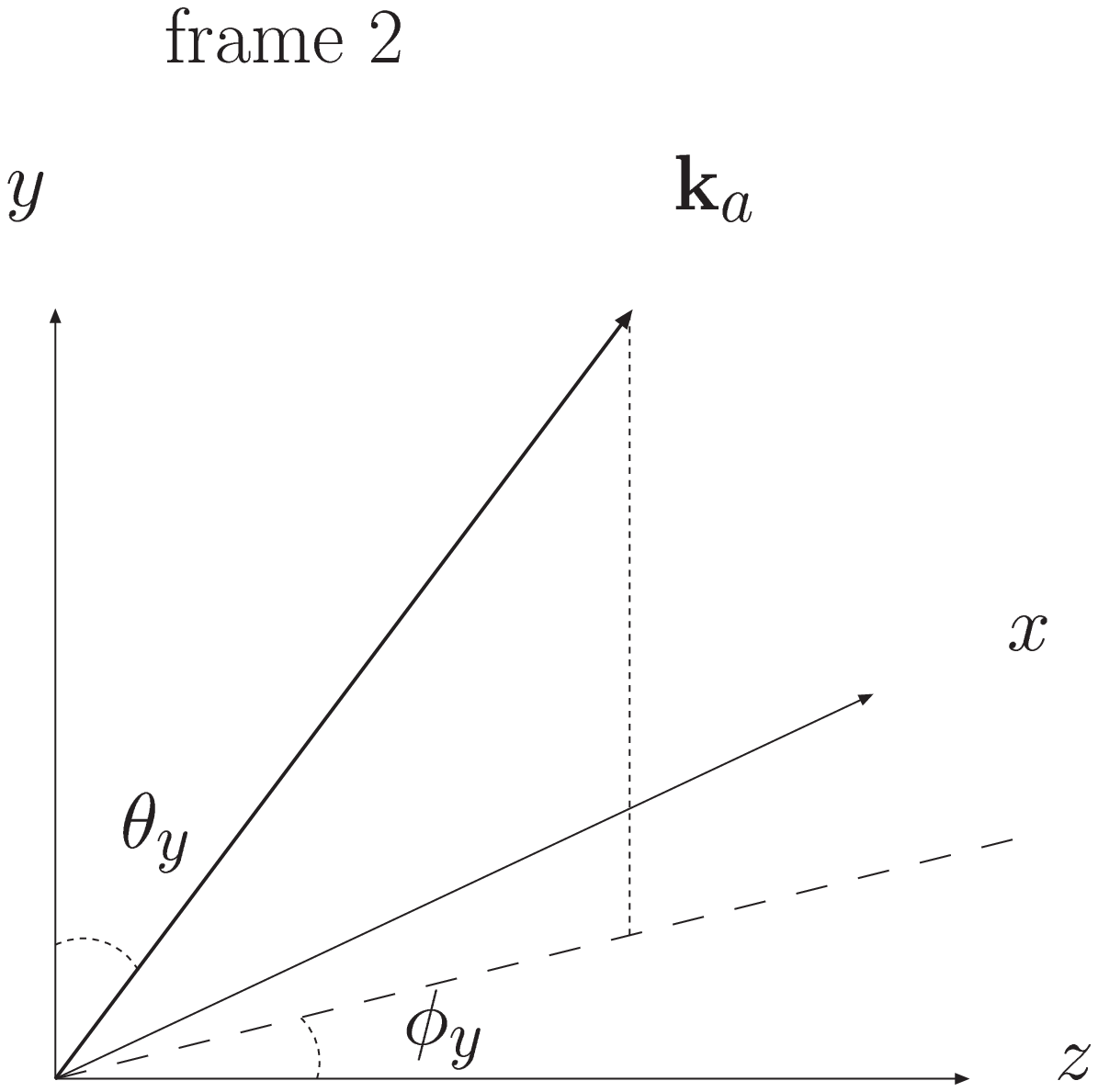}
    \caption{The polar coordinate of frame 2. The North pole is $y$ direction.}
  \label{fig:frame2thetaphi-y.eps}
   \end{minipage}
\end{figure}

As we shall see, angular distribution of the particle $a$ in $\tau\to\nu_\tau+a+\mathrm{anything}$ will give us information about $a_-$, $b_-$, $c_-$, $d_-$, $e_-$, $f_+$ and $g_+$.
     The general formula (\ref{sum pm s ^pm}) contains all the information. 
      However, it is too complex to analyze it here. 
    So, we integrate about $d\Omega$ to simplify the formula.
     Furthermore, we introduce three formulae, which are the general formulae integrated about three kinds of azimuthal angles of the momentum $\mathbf{k}_a$ as explained in Figs. \ref{fig:frame2thetaphi-z.eps}-\ref{fig:frame2thetaphi-y.eps} to simplify the formula, respectively.
    Thanks to this prescription, we obtain three simple formulae which are convenient to analyze here. 

  Consider the double differential distribution, where $d^3k_a$ is replaced by $dk_a d\cos\theta_z d\phi_z$: 
\begin{align} \begin{split}\label{generalpol}
&\frac{d\sigma}
{   dx_1 dx_2 d\Omega_\tau  d\psi \ dy_a d\phi_{az} d\cos\theta_{az} }   
 \\&=
   Br(\tau \to \mu \nu \bar\nu) Br\bigl(\tau^- \to \nu_\tau+a+\mathrm{anything}\bigr)
\frac{  \alpha^2 \beta y_a^2 }{\pi^2 q^2  \lambda_a}   
\\ &\hspace{1em}\times
\biggl[ G_0(x_1,x_2) G_1^a(y_a) \bigl(2+\frac{1}{\gamma^2}\bigr)  
\\  &\hspace{2.5em} -\sum_i G_i^s(x_1,x_2)G_2^a(y_a)
\Bigl\{  \bigl(2-\frac{1}{\gamma^2}\bigr) \hat  k_{az}  P_{iz}-\beta^2     \hat k_{ay} P_{iy}  
           +\bigl(1+\frac{1}{\gamma^2}\bigr)   \hat k_{ax}  P_{ix}
\Bigr\}
\biggr],
\end{split} \end{align}
where $y_a=2E_a/m_\tau$.
$\sum_i G_i^s(x_1,x_2)P_{iz}$, $\sum_i G_i^s(x_1,x_2)P_{ix}$, $\sum_i G_i^s(x_1,x_2)P_{iy}$ can be obtained by considering three single differential distributions as follows.

Integrate $d\phi_{az}$, then  
\begin{align} \begin{split}
&\frac{d\sigma}
{   dx_1 dx_2 d\Omega_\tau  d\psi \ dy_a   d\cos\theta_{az} }   
 \\&=
 2\pi  Br(\tau \to \mu \nu \bar\nu) Br\bigl(\tau^- \to \nu_\tau+a+\mathrm{anything}\bigr)
\frac{  \alpha^2 \beta y_a^2 }{\pi^2 q^2  \lambda_a}   
\\ &\hspace{1em}\times
\biggl[ G_0(x_1,x_2) G_1^a(y_a) \bigl(2+\frac{1}{\gamma^2}\bigr)  
\\  &\hspace{2.5em} -\sum_i G_i^s(x_1,x_2)G_2^a(y_a)
  \bigl(2-\frac{1}{\gamma^2}\bigr) \hat  k_{az}  P_{iz}  
\biggr].
\end{split} \end{align}
And using the equation
\begin{align} \begin{split}
&\frac{d\sigma}
{   dx_1 dx_2 d\Omega_\tau  d\psi \ dy_a    }   
 \\&=
4\pi  Br(\tau \to \mu \nu \bar\nu) Br\bigl(\tau^- \to \nu_\tau+a+\mathrm{anything}\bigr)
\frac{  \alpha^2 \beta y_a^2 }{\pi^2 q^2  \lambda_a}   
\\ &\hspace{1em}\times
 G_0(x_1,x_2) G_1^a(y_a) \bigl(2+\frac{1}{\gamma^2}\bigr)  
,
\end{split} \end{align}
then
\begin{align} \begin{split}\label{Eq a}
&\frac{d\sigma}
{   dx_1 dx_2 d\Omega_\tau  d\psi \ dy_a   d\cos\theta_{az} }   -\frac{1}{2}\frac{d\sigma}
{   dx_1 dx_2 d\Omega_\tau  d\psi \ dy_a    }
 \\&=
 -2\pi  Br(\tau \to \mu \nu \bar\nu) Br\bigl(\tau^- \to \nu_\tau+a+\mathrm{anything}\bigr)
\frac{  \alpha^2 \beta y_a^2 }{\pi^2 q^2  \lambda_a}   
\\ &\hspace{1em}\times
G_2^a(y_a)    \bigl(2-\frac{1}{\gamma^2}\bigr) \hat  k_{az}
 \sum_i G_i^s(x_1,x_2)
   P_{iz}  .
\end{split} \end{align}

Similarly, Taking the coordinates as Figs. \ref{fig:frame2thetaphi-x.eps} and \ref{fig:frame2thetaphi-y.eps}, we have the equations 
\begin{align} \begin{split}
&\frac{d\sigma}
{   dx_1 dx_2 d\Omega_\tau  d\psi \ dy_a   d\cos\theta_{ax} }  -\frac{1}{2}\frac{d\sigma}
{   dx_1 dx_2 d\Omega_\tau  d\psi \ dy_a    }    
 \\&=
- 2\pi  Br(\tau \to \mu \nu \bar\nu) Br\bigl(\tau^- \to \nu_\tau+a+\mathrm{anything}\bigr)
\frac{  \alpha^2 \beta y_a^2 }{\pi^2 q^2  \lambda_a}   
\\ &\hspace{1em}\times
G_2^a(y_a)
\bigl(1+\frac{1}{\gamma^2}\bigr)   \hat k_{ax} \sum_i G_i^s(x_1,x_2)  P_{ix}
\end{split} \end{align}
and
\begin{align} \begin{split}
&\frac{d\sigma}
{   dx_1 dx_2 d\Omega_\tau  d\psi \ dy_a   d\cos\theta_{ay} }    -\frac{1}{2}\frac{d\sigma}
{   dx_1 dx_2 d\Omega_\tau  d\psi \ dy_a    }    
 \\&=
  2\pi Br(\tau \to \mu \nu \bar\nu) Br\bigl(\tau^- \to \nu_\tau+a+\mathrm{anything}\bigr)
\frac{  \alpha^2 \beta y_a^2 }{\pi^2 q^2  \lambda_a}   
\\ &\hspace{1em}\times
G_2^a(y_a)
  \beta^2     \hat k_{ay}  \sum_i G_i^s(x_1,x_2)  P_{iy}  
,
\end{split} \end{align}
respectively.
%
\\
From the above three equations, we pull out the quantities,
\begin{align} \begin{split}
 &\sum_i P_{ix}G_i^s(x_1,x_2)  
 \\&\sum_i P_{iy}G_i^s(x_1,x_2)  
 \\&\sum_i P_{iz} G_i^s(x_1,x_2). 
\end{split} \end{align}

First, using these quantities, we analyze $ G_1^s(x_1,x_2)$ and $ G_2^s(x_1,x_2) $ contemporary.
  $G_{\hat{\mathbf{p}}_3}^s(x_1,x_2)$ which is the $\hat{\mathbf{p}}_3 $ component of $  \sum_i    \mathbf{P}_i   G_i^s(x_1,x_2)$ is defined as
    \begin{align} \begin{split}
  G_{\hat{\mathbf{p}}_3}^s(x_1,x_2)
    = \hat{\mathbf{p}}_3 \cdot \sum_i    \mathbf{P}_i   G_i^s(x_1,x_2)
  =  \hat{\mathbf{p}}_3 \cdot \hat{\mathbf{p}}_1   G_1^s(x_1,x_2) +\hat{\mathbf{p}}_3 \cdot \hat{\mathbf{p}}_2 G_2^s(x_1,x_2)
  \\=\hat{p}_{3x}\sum_i P_{ix}G_i^s(x_1,x_2)  
 +\hat{p}_{3y}\sum_i P_{iy}G_i^s(x_1,x_2)  
 +\hat{p}_{3z}\sum_i P_{iz} G_i^s(x_1,x_2),
    \end{split} \end{align}
  where    
\begin{align} \begin{split}\label{Eq d}
  \hat{\mathbf{p}}_3 \cdot \hat{\mathbf{p}}_1&=1-2\frac{1-x_2}{x_1 x_3}
\\\hat{\mathbf{p}}_3 \cdot \hat{\mathbf{p}}_2&=1-2\frac{1-x_1}{x_2 x_3}.
\end{split} \end{align}

 Integrating over $x_2$ and  defining the parameters 
\begin{align} \begin{split}\label{Eq -}
a_-&=\bigl(\frac{|g_1|^2}{16}+|g_3|^2\bigr)-\bigl( \frac{|g_2|^2}{16}+|g_4|^2\bigr)    \\
b_-&=|g_5|^2-|g_6|^2      \\
c_-&= |eA_R|^2 - |eA_L|^2 \\
d_-&=   -\bigl( Re[g_3 e A_L^*]  -Re[g_4 e A_R^*]\bigr)  \\
e_-&= -\bigl(Re[g_6 e A_R^*]  - Re[g_5 e A_L^*]\bigr),    
\end{split} \end{align}
 $ G_{\hat{\mathbf{p}}_3}^s(x_1,x_2)$ becomes
\begin{align} \begin{split}
 G_{\hat{\mathbf{p}}_3}^s(x_1)&= \in {1-x_1} {x_1} dx_2 G_{\hat{\mathbf{p}}_3}^s(x_1,x_2) 
   \\ =&  
 \frac{1}{6}\biggl[
4(2x_1-1)\Bigl\{a_-(2x_1-1)(5-4x_1)+12d_-(2x_1-1)+3e_-(3-2x_1) \Bigr\} 
 \\&\hspace*{1em} +8c_-   \Bigl\{2(2x_1-1)\frac{(x_1^2-13x_1+13)}{(1-x_1)}
  \\&\hspace*{4.5em}-3(2x_1^2-2x_1+1)\log\bigl[\frac{1-x_1}{x_1}\bigr]   
  +24\log\bigl[2(1-x_1)\bigr]\Bigr\}     
  \\&\hspace*{1em}  +b_-   \Bigl\{ (8x_1^2-32x_1+23)(1-2x_1)-24 (1-x_1)^2\log\bigl[2(1-x_1)\bigr] \Bigr\} \biggr].   \\
\end{split} \end{align}

To determine the value of coefficient $c_- $, we define the function 
\begin{eqnarray}
F_5(x_1 )
=
\frac{3(1-x_1)}{8 }  G_{\hat{\mathbf{p}}_3}^s(x_1).
\end{eqnarray}
The value of $F_5$ where $x_1=1$ is
\begin{eqnarray}
F_5(x_1)\bigr|_{x_1=1}
= 
    c_-. 
\end{eqnarray} 
So we can determine the value of $c_-$.

Then, to determine the values of $a_- $, $b_- $,  $d_- $ and $e_- $,  we define another function  subtracting  the term of coefficient $c_- $, 
\begin{align} \begin{split}
   F_6(x_1 )    
=  
 6\left\{G_{\hat{\mathbf{p}}_3}^s(x_1)-(c_-\ \mathrm{term})\right\},   
\end{split} \end{align}
where 
\begin{align} \begin{split}
(c_-\ \mathrm{term})= 8c_-   \Bigl\{&2(2x_1-1)\frac{(x_1^2-13x_1+13)}{(1-x_1)}
\\ &-3(2x_1^2-2x_1+1)\log\bigl[\frac{1-x_1}{x_1}\bigr]   
   +24\log\bigl[2(1-x_1)\bigr]\Bigr\}.  
\end{split} \end{align}

 We'll find that the 4 aspects of this function lead to determination of  parameters  $a_- $, $b_- $,  $d_- $ and $e_- $.   
The value of $F_6$ where $x_1=1$ is
\begin{eqnarray}
F_{6a}=F_6(x_1 ) \bigr|_{x_1=1}=4a_-+b_-+48d_-+12e_-.
\end{eqnarray}
The gradient of $F_6$ where $x_1=1$ leads
\begin{eqnarray}
F_{6b}=\frac{1}{6}\frac{d}{dx_1}F_6(x_1 )\Bigr|_{x_1=1}= 3b_-+32d_-.
\end{eqnarray}
The gradient of $F_6$ where $x_1=1/2$ leads
\begin{eqnarray}
F_{6c}=\frac{1}{6}\frac{d}{dx_1}F_6(x_1 )\Bigr|_{x_1=\frac{1}{2}}=-b_-+8e_-.
\end{eqnarray}
The integration value of $F_6$ leads
\begin{align} \begin{split}
F_{6d}= 6\in {\frac{1}{2}} 1 dx_1 F_6(x_1 ) 
=6a_--b_-+48d_-+24e_-.
\end{split} \end{align}
From previous four equations, we can determine the parameters  $a_- $, $b_- $,  $d_- $ and $e_- $: 
\begin{align}
a_-&=    \frac{1}{8}(-10F_{6a}+3F_{6b}-9F_{6c}+8F_{6d}),  \\
b_-&= \frac{1}{2}( -6F_{6a}+3F_{6b}-3F_{6c}+4 F_{6d}),  \\
d_-&=\frac{1}{64}(18F_{6a}-7F_{6b}+9F_{6c}-12F_{6d}),  \\
e_-&=\frac{1}{16}(-6F_{6a}+3F_{6b}-F_{6c}+4F_{6d}).
 \end{align}

Next, we analyze  $G_3^s(x_1,x_2)$ using the relation 
\begin{equation}
G_3^s(x_1,x_2)=
\frac{(\hat{ \mathbf{ p}}_1 \times \hat{ \mathbf{ p}}_2)}{|\hat{ \mathbf{ p}}_1 \times \hat{ \mathbf{ p}}_2|^2}\cdot\sum_i \mathbf{P}_{i}G_1^s(x_1,x_2).  
\end{equation}
Integrating it over $x_2$, we define the quantity $G_3^s(x_1)$ as 
\begin{align} \begin{split}
G_3^s(x_1)&= \in {1-x_1} {x_1}  dx_2 \ G_3^s(x_1,x_2)
 \\&=
   \frac{ x_1 }{3 (1- x_1 )} 
    \biggl[4 f_+ \Bigl\{ (2  x_1-1)(2  x_1 ^2 -2x_1-1)+3 ( x_1 -1) x_1 \log \bigl[\frac{1-x_1}{x_1}\bigr]
    \Bigr\} 
    \\&\hspace{6em} +g_+  \Bigl\{  (2  x_1-1)  (4 x_1^2-10 x_1 +7) +6( x_1 -1)^2 \log \bigl[\frac{1-x_1}{x_1}\bigr]\Bigr\}
   \biggr],
   \end{split} \end{align}
     where
\begin{align} \begin{split}\label{Eq im}
 f_+&= -\bigl(Im[g_3 e A_L^*] + Im[g_4 e A_R^*]\bigr) 
\\g_+&= -\bigl( Im[g_6 e A_R^*]+Im[g_5 e A_L^*]\bigr).
\end{split} \end{align}
Here, we define the function
\begin{align} \begin{split}
F_7(x_1)=G_3^s(x_1)    \frac{6(1-x_1)}{(1-2x_1)^2}.
\end{split} \end{align}
Thus we determine $f_+$ and $g_+$ from 
\begin{align} \begin{split}
F_7(x_1)\Bigr|_{x_1=1}&=-8f_+ +2g_+\\
F_7(x_1)\Bigr|_{x_1=\frac{1}{2}}&=3g_+.
\end{split} \end{align}

             So we can determine the values of $a_- $, $b_-$, $c_- $, $d_- $, $e_- $,  $f_+$ and $g_+$ separately.
        Adding this result to the result of the previous section i.e. $a_+ $, $b_+$, $c_+ $, $d_+ $ and $e_+ $, we can determine the values of $|g_1|^2/16+|g_3|^2$, $|g_2|^2/16+|g_4|^2$, $|g_5|^2$, $|g_6|^2$, $|A_R|^2$, $|A_L|^2$, $Re[g_3  A_L^*]$,   $Re[g_4  A_R^*]$, $Re[g_5  A_L^*]$, $Re[g_6  A_R^*]$, $Im[g_3  A_L^*] + Im[g_4  A_R^*]$ and $Im[g_5  A_L^*] + Im[g_6  A_R^*]$, separately. 
            Moreover, in some suitable cases, we can determine   $|g_1|$ and  $|g_3|$ ( also $|g_2|$ and  $|g_4|$), separately.
        To separate these are the main purpose of following subsection. 

\section{Physics Implication from $G_i^s(x_3)$ Distribution of $\mu_3$ }\label{physics implication}

In sections \ref{sec 3} and \ref{Energymu3}, we studied that $a_+$, $b_+$, $c_+$, $d_+$ and $e_+$ are determined by the energy distribution.
In section \ref{sec 5}, we also studied that $a_-$, $b_-$, $c_-$, $d_-$, $e_-$, $f_+$ and $g_+$ are determined by the energy distribution and the angular distribution.

Here, we study some special cases using these parameters.
%
%
Using Eq. (\ref{rtheta}),  each parameter defined in (\ref{Eq -}) and (\ref{Eq im}) are represented as
\begin{align} \begin{split}
a_-&=\frac{1}{16}(r_1^2-r_2^2)+r_3^2-r_4^2  \\
b_-&=r_5^2-r_6^2  \\
 c_-& =r_R^2-r_L^2 \\
 d_-&=-r_3 r_L\cos(\theta_3-\theta_L)+ r_4 r_R\cos(\theta_4-\theta_R)\\
 e_-&=- r_6 r_R\cos(\theta_6-\theta_R)+r_5 r_L\cos(\theta_5-\theta_L)\\
  f_+&=-r_3 r_L\sin(\theta_3-\theta_L)- r_4 r_R\sin(\theta_4-\theta_R)\\
 g_+&=- r_6 r_R\sin(\theta_6-\theta_R)-r_5 r_L\sin(\theta_5-\theta_L)
.
\end{split} \end{align}

 \subsection{A sufficient condition for existence of scalar and/or pseudo scalar current }\label{6.9}
 We make three types of relations which may reveal the existence of scalar and/or pseudo scalar current in this and following two subsections. 

First, in this subsection, we give a sufficient condition for existence of $g_1$ and/or $g_2$.
  Using the fact that
\begin{align} \begin{split}
\left(r_3r_4\sin(\theta_3-\theta_L)\sin(\theta_4-\theta_R)+\sqrt{\frac{r_1^2}{16}+r_3^2}\sqrt{\frac{r_2^2}{16}+r_4^2}\right)\ge 0
\end{split} \end{align}
and the relation
\begin{align} \begin{split}\label{Eq d2}
&f_+^2 +\frac{d_+^2 + d_-^2}{2}-\frac{a_+c_+-a_-c_-}{2}
\\&=r_3^2r_L^2+r_4^2r_R^2+2r_3r_Lr_4r_R\sin(\theta_3-\theta_L)\sin(\theta_4-\theta_R)-\frac{a_+c_+-a_-c_-}{2}
\\&=-\frac{r_1^2}{16}\frac{c_+-c_-}{2}-\frac{r_2^2}{16}\frac{c_++c_-}{2}+2r_3r_Lr_4r_R\sin(\theta_3-\theta_L)\sin(\theta_4-\theta_R),
\end{split} \end{align} 
we can make a relation
\begin{align} \begin{split}
&\frac{|g_1|^2|eA_L|^2+|g_2|^2|eA_R|^2}{16}\\
&\ge
\frac{r_1^2r_L^2+r_2^2r_R^2}{16}-2r_Lr_R\left(r_3r_4\sin(\theta_3-\theta_L)\sin(\theta_4-\theta_R)+\sqrt{\frac{r_1^2}{16}+r_3^2}\sqrt{\frac{r_2^2}{16}+r_4^2}\right)\\
&=\frac{a_+c_+-a_-c_-}{2}
-\frac{1}{2}\sqrt{(a_+^2-a_-^2)(c_+^2-c_-^2)}  -f_+^2 -\frac{d_+^2 + d_-^2}{2}.
\end{split} \end{align}

If
\begin{align} \begin{split}
\frac{a_+c_+-a_-c_-}{2}
-\frac{1}{2}\sqrt{(a_+^2-a_-^2)(c_+^2-c_-^2)} -f_+^2 -\frac{d_+^2 + d_-^2}{2}
>0,
\end{split} \end{align}
 then
 \begin{align} \begin{split}
|g_1|^2|eA_L|^2+|g_2|^2|eA_R|^2
>0.
\end{split} \end{align}
So, in this case, we can get the result that $g_1 $ and/or $ g_2 \not=0$.

\subsection{What can we say about $g_1$, $g_2$ and/or CP Violation}\label{6.10}
We now give relations which are convenience for determining the existence of scalar and/or pseudo scalar currents and/or CP violation of Lagrangian.
One of them is
\begin{align} \begin{split}
\frac{a_+ + a_-}{2}-(\frac{d_+ + d_-}{2})^2\frac{2}{c_+ - c_-}
=r_3^2\sin^2(\theta_3-\theta_L)+\frac{r_1^2}{16}  \ge 0.
\end{split} \end{align}
 If the left hand side becomes zero and $a_++a_-\not=0$, then $\sin(\theta_3-\theta_L)=g_1=0 $ and $(a_++a_-)/2=|g_3|^2$.
 On the other hand, if the left hand side is larger than zero, at least, $|g_3|\sin(\theta_3-\theta_L)\not=0$ or $g_1\not=0 $.
$|g_3|\sin(\theta_3-\theta_L)\not=0$ means CP violation. 
$g_1\not=0 $ means that the scalar and/or pseudo scalar current exists.
 Similarly, we make the relation,
 \begin{align} \begin{split}
\frac{a_+ - a_-}{2}-(\frac{d_+ - d_-}{2})^2\frac{2}{c_+ + c_-}
=r_4^2\sin^2(\theta_4-\theta_R)+\frac{r_2^2}{16}  \ge 0.
\end{split} \end{align}
If the left hand side becomes zero and $a_+-a_-\not=0$, then $\sin(\theta_4-\theta_R)=g_2=0 $ and $(a_+-a_-)/2=|g_4|^2$.
 On the other hand, if the left hand side is larger than zero, at least, $|g_4|\sin(\theta_4-\theta_R)\not=0$ or $g_2\not=0 $.
$|g_4|\sin(\theta_4-\theta_R)\not=0$ means CP violation. 
$g_2\not=0 $ means the that scalar and/or pseudo scalar current exists.

\subsection{Existence of $g_1$ and/or $g_2$, or the Values of $Im[g_3 e A_L^*]$ and $Im[g_4 e A_R^*]$}\label{6.11}
Finally, we have two more relations. First one is
\begin{align} \begin{split}\label{Eq e}
&\sqrt{\frac{a_+ + a_-}{2}\frac{c_+ - c_-}{2}-(\frac{d_+ + d_-}{2})^2}
+\sqrt{\frac{a_+ - a_-}{2}\frac{c_+ + c_-}{2}-(\frac{d_+ - d_-}{2})^2}\\
&=r_L\sqrt{r_3^2\sin^2(\theta_3-\theta_L)+\frac{r_1^2}{16}} 
+r_R\sqrt{r_4^2\sin^2(\theta_4-\theta_R)+\frac{r_2^2}{16}}\\
&\ge r_3r_L|\sin(\theta_3-\theta_L)| 
+    r_4r_R|\sin(\theta_4-\theta_R)|\\
&  \ge |f_+|.
\end{split} \end{align}

It becomes an equality when 
\begin{eqnarray}\label{Eq c}
 \left\{
  \begin{array}{l}
\ g_1=g_2=0      \\
\ Im[g_3 e A_L^*] Im[g_4 e A_R^*] \ge 0.       \\
  \end{array}
\right.
\end{eqnarray}
In that case, from the relation (\ref{Eq e}) and the conditions (\ref{Eq c}), $Im[g_3 e A_L^*]$ and $ Im[g_4 e A_R^*]$ are determined as
\begin{align} \begin{split}
Im[g_3 e A_L^*]&=-\frac{f_+}{|f_+|}\sqrt{\frac{a_+ + a_-}{2}\frac{c_+ - c_-}{2}-(\frac{d_+ + d_-}{2})^2}\\
Im[g_4 e A_R^*]&=-\frac{f_+}{|f_+|}\sqrt{\frac{a_+ - a_-}{2}\frac{c_+ + c_-}{2}-(\frac{d_+ - d_-}{2})^2}.
\end{split} \end{align}
If $f_+=0$, it means $Im[g_3 e A_L^*]= Im[g_4 e A_R^*]=0$ since $ Im[g_3 e A_L^*] Im[g_4 e A_R^*] \ge 0$.

We derive another relation from (\ref{Eq d2})
\begin{align} \begin{split}
2Im[g_3 e A_L^*] Im[g_4 e A_R^*]
&
=  f_+^2
-\frac{a_+c_+- a_-c_-}{2}+\frac{d_+^2 + d_-^2}{2}\\
&+\frac{1}{16}( |g_1|^2 |e A_L|^2 +|g_2|^2  |e A_R|^2 ) 
\\&\ge\frac{1}{2}\left\{2f_+^2-(a_+c_+- a_-c_-)+(d_+^2 + d_-^2)\right\}.
\end{split} \end{align}
When the equation of (\ref{Eq e}) is not an equality and $ 2f_+^2-(a_+c_+- a_-c_-)+(d_+^2 + d_-^2) \ge 0$
, then we can derive, using (\ref{Eq c}), that  $g_1 \not =0 $ and/or $ g_2\not =0 $.
Furthermore, if $ 2f_+^2-(a_+c_+- a_-c_-)+(d_+^2 + d_-^2) > 0$
, then $Im[g_3 e A_L^*]\not=0$ and $Im[g_4 e A_R^*]\not=0$ since $Im[g_3 e A_L^*] Im[g_4 e A_R^*]>0$.
In that case, CP symmetry is violated and the sign of $Im[g_3 e A_L^*]$ and $ Im[g_4 e A_R^*]$ is the same as that of $-f_+$.

\subsection{$Im[  g_5 e A_L^*]$ and $Im[  g_6 e A_R^*]$}\label{6.12}

Here, we give the method to determine $Im[  g_5 e A_L^*]$ and $Im[  g_6 e A_R^*]$.

Substituting the relations
\begin{align} \begin{split}
Im[  g_5 e A_L^*]^2&= |g_5|^2 |e A_L|^2 -Re[  g_5 e A_L^*]^2   
\\
Im[  g_6 e A_R^*]^2&= |g_6|^2 |e A_R|^2 -Re[  g_6 e A_R^*]^2   
\end{split} \end{align}
to
\begin{align} \begin{split}
Im[  g_5 e A_L^*]^2
&=  \bigl(g_+ +Im[  g_6 eA_R^*]\bigr)^2
\\&=Im[  g_6 eA_R^*]^2 +g_+^2+2g_+ Im[  g_6 eA_R^*],
\end{split} \end{align}
the imaginary part of $ g_6 e A_R^*$ is represented only by the observables as
\begin{align} \begin{split}
  Im[  g_6 eA_R^*] 
&=\frac{-Im[  g_6 eA_R^*]^2 -g_+^2+Im[  g_5 eA_L^*]^2}{2g_+}
\\&=-\frac{|g_6|^2 |eA_R|^2 -Re[  g_6 eA_R^*]^2 +g_+^2-|g_5|^2 |eA_L|^2 +Re[  g_5 eA_L^*]^2}{2g_+}
\\&=-\frac{b_+c_--b_-c_+-2e_+e_-+2g_+^2}{4g_+}.
\end{split} \end{align}
So, if $g_+ \not =0 $, $Im[  g_6 e A_R^*]$ can be determined independently. Similarly, $Im[  g_5 e A_L^*]$ is represented as
\begin{align} \begin{split}
  Im[  g_5 e A_L^*] 
&=\frac{Im[  g_5 e A_L^*]^2 +g_+^2-Im[  g_6 e A_R^*]^2}{2g_+}
\\&=\frac{|g_5|^2 |e A_L|^2 -Re[  g_5 A_L^*]^2 +g_+^2-|g_6|^2 |e A_R|^2 +Re[  g_6 e A_R^*]^2}{2g_+}
\\&=-\frac{b_-c_+-b_+c_-+2e_+e_-+2g_+^2}{4g_+},
\end{split} \end{align}
if $g_+ \not =0 $.

Even if $g_+  =0 $,
\begin{align} \begin{split}
\frac{c_++c_-}{2}\frac{b_+-b_-}{2}-\left(\frac{e_++e_-}{2}\right)^2=Im[g_6eA_R^*]^2
\end{split} \end{align}
and
\begin{align} \begin{split}
\frac{c_+-c_-}{2}\frac{b_++b_-}{2}-\left(\frac{e_+-e_-}{2}\right)^2=Im[g_5eA_L^*]^2.
\end{split} \end{align}
So we can determine the absolute values of $Im[g_5eA_L^*]$ and $Im[g_6eA_R^*]$.

\subsection{What can we say if one of $A_R, A_L, g_3, g_4=0$}\label{6.13}

If we restrict that $A_R=0$ from other experiments or some specific models, then 
\begin{align} \begin{split}
f_+&=-r_3r_L\sin(\theta_3-\theta_L) \\
d_+&=-r_3r_L \cos(\theta_3 -\theta_L)   \\
c_+&=r_L^2.
\end{split} \end{align}
So,
\begin{align} \begin{split}
\frac{d_+^2+f_+^2}{c_+}
&=r_3^2=|g_3|^2
\\
\frac{a_++a_-}{2}-\frac{d_+^2+f_+^2}{c_+}
&=\frac{r_1^2}{16}=\frac{|g_1|^2}{16}
\\c_+&=|eA_L|^2.
\end{split} \end{align}
In this case, $|g_3|$ and $|g_1|$ are determined independently. 
Similarly, if $A_L=0$,
\begin{align} \begin{split}
\frac{d_+^2+f_+^2}{c_+}
& =|g_4|^2
\\
\frac{a_+-a_-}{2}-\frac{d_+^2+f_+^2}{c_+}
& =\frac{|g_2|^2}{16}
\\c_+&=|eA_R|^2;
\end{split} \end{align}
if $g_4=0$, 
\begin{align} \begin{split}
(d_+^2+f_+^2)\frac{2}{c_+-c_-}& =|g_3|^2
\\\frac{a_++a_-}{2}-(d_+^2+f_+^2)\frac{2}{c_+-c_-}& =\frac{|g_1|^2}{16}
\\
\frac{a_+-a_-}{2}&=\frac{|g_2|^2}{16};
\end{split} \end{align}
and if $g_3=0$,
\begin{align} \begin{split}
(d_+^2+f_+^2)\frac{2}{c_++c_-}&=|g_4|^2
\\\frac{a_+-a_-}{2}-(d_+^2+f_+^2)\frac{2}{c_++c_-}&=\frac{|g_2|^2}{16}
\\\frac{a_++a_-}{2}&=\frac{|g_1|^2}{16}.
\end{split} \end{align}

\subsection{CP violation}\label{6.14}

The master formula (\ref{sum pm s ^pm}) contains a part,
\begin{align} \begin{split}\label{no sum 3}
&  \bigl(1+\cos^2\eta -\frac{\sin ^2 \eta}{\gamma^2}\bigr) \hat  k_{az}  P_{3z}-\beta^2 \sin^2\eta \  \hat k_{ay} P_{3y}  
\\  &\hspace{6em}+\bigl(1+\frac{1}{\gamma^2}\bigr) \sin^2\eta \  \hat k_{ax}  P_{3x} - \frac{\sin 2 \eta}{\gamma} (\hat  k_{ax}  P_{3z}+\hat  k_{az}  P_{3x}).
\end{split} \end{align}
In this part, each term is proportional to
\begin{align}\label{vectors}
 \hat  k_{az}  (\hat{\mathbf{p}}_1\times \hat{\mathbf{p}}_1)_z,&&
   \hat k_{ay}  (\hat{\mathbf{p}}_1\times \hat{\mathbf{p}}_1)_y,&&  
  \hat k_{ax}   (\hat{\mathbf{p}}_1\times \hat{\mathbf{p}}_1)_x,&&
  \hat  k_{ax}   (\hat{\mathbf{p}}_1\times \hat{\mathbf{p}}_1)_z+\hat  k_{az}   (\hat{\mathbf{p}}_1\times \hat{\mathbf{p}}_1)_x.
\end{align}

Supposing CPT theorem, time reversal is equivalent to CP transformation.
In time reversal, momentum flips their directions.
In fact,
\begin{align}
\hat{\mathbf{k}}_a \to -\hat{\mathbf{k}}_a ,&&
\hat{\mathbf{p}}_1 \to -\hat{\mathbf{p}}_1,&& 
\hat{\mathbf{p}}_2 \to -\hat{\mathbf{p}}_2. 
\end{align}
In this transformation,
 (\ref{vectors}) flip their signs.
 This means (\ref{no sum 3}) part in (\ref{sum pm s ^pm}) flips its sign.
This part proportions to $g_+$ or $f_+$.
These coefficients are the imaginary parts of the interactions.
So, if they do not vanish, then CP symmetry is violated.

\subsection{$A_L,A_R\gg \mathrm{others}$}

If $\alpha Br(\tau\to\mu\gamma)$ is large compared to 4-fermi sector of $Br(\tau\to 3\mu)$, here $\alpha$ is the fine structure constant, it may be difficult to determine $|g_1|$ to $|g_6|$ directly from $\tau\to 3\mu$.
It may be still possible to get a bound for these coupling constant from interference effects as follows.
\begin{align} \begin{split}
|g_3|^2=r_3^2\ge r_3^2 \cos^2(\theta_3-\theta_L)=\frac{(d_++d_-)^2}{2(c_+-c_-)}
\end{split} \end{align}
\begin{align} \begin{split}
|g_4|^2=r_4^2\ge r_4^2 \cos^2(\theta_4-\theta_R)=\frac{(d_+-d_-)^2}{2(c_++c_-)}
\end{split} \end{align}

\begin{align} \begin{split}
|g_5|^2=r_5^2\ge r_5^2 \cos^2(\theta_5-\theta_L)=\frac{(e_+-e_-)^2}{2(c_+-c_-)}
\end{split} \end{align}
\begin{align} \begin{split}
|g_6|^2=r_6^2\ge r_6^2 \cos^2(\theta_6-\theta_R)=\frac{(e_++e_-)^2}{2(c_++c_-)}.
\end{split} \end{align}

\subsection{$A_L,A_R\ll \mathrm{others}$}\label{a<o}

If $\alpha Br(\tau\to\mu\gamma)$ is small compared to $Br(\tau\to 3\mu)$, we can still get some lower bounds from the amplitudes of $\tau\to\mu\gamma$ decay as follows.
\begin{align} \begin{split}
|eA_L|^2=r_L^2\ge \frac{r_3^2r_L^2\cos^2(\theta_3-\theta_L)}{\frac{r_1^2}{16}+r_3^2}=\frac{(d_++d_-)^2}{2(a_++a_-)}
\end{split} \end{align}
\begin{align} \begin{split}
|eA_L|^2=r_L^2\ge \frac{r_5^2r_L^2\cos^2(\theta_5-\theta_L)}{r_5^2}=\frac{(e_+-e_-)^2}{2(b_++b_-)}
\end{split} \end{align}
\begin{align} \begin{split}
|eA_R|^2=r_R^2\ge \frac{r_4^2r_R^2\cos^2(\theta_4-\theta_R)}{\frac{r_2^2}{16}+r_4^2}=\frac{(d_+-d_-)^2}{2(a_+-a_-)}
\end{split} \end{align}
\begin{align} \begin{split}
|eA_R|^2=r_R^2\ge \frac{r_6^2r_R^2\cos^2(\theta_6-\theta_R)}{r_6^2}=\frac{(e_++e_-)^2}{2(b_+-b_-)}
\end{split} \end{align}

\section{Concluding Remarks}\label{conclud}
 
Neutrino oscillation suggests that, flavor quantum number isn't conserved not only in quark sector but also in neutrino sector.
KM ansatz implies that if neutrino sector violate the flavor quantum number, charged lepton sector also do it.
In the Standard Model, this violation is too small to determine in any designed futur experiments.
However, in some new models, it is suggested that this violation is going to be determine.
If $\tau\to 3\mu$ event is detected, there are many models which are suitable to the first experimental result.  
However, at least, all models without one model is not true.  
So, if $\tau\to 3\mu$ event is detected, our analysis must be necessary to figure out if one model is allowed or forbidden.

   We assumed only Lorentz and gauge invariance of the Lagrangian and 
   locality of the action.
 So, if we cannot fit the data to the differential cross section, it means violation of Lorentz or gauge invariance. 

       From energy distributions $ (|g_1|^2/16+|g_3|^2 )+( |g_2|^2/16+|g_4|^2)$, $|g_5|^2+|g_6|^2$, $|A_R|^2 +|A_L|^2$, $Re[g_4 A_R^*]+Re[g_3 A_L^*]$ and $Re[g_6 A_R^*]+Re[g_5 A_L^*]$   can be determined.
       Using the angular distribution of decaying products of $\tau^+$ and $\tau^-$, we can determine $ (|g_1|^2/16+|g_3|^2 )$, $( |g_2|^2/16+|g_4|^2)$, $|g_5|$, $|g_6|$, $|A_R| $, $|A_L|$, $Re[g_4 A_R^*]  $, $Re[g_3 A_L^*]$, $Re[g_6 A_R^*] $, $Re[g_5 A_L^*]$, $Im[g_4 A_R^*] + Im[g_3 A_L^*]$ and $Im[g_6 A_R^*]  + Im[g_5 A_L^*]$ independently. 
      We can determine the argument of relative phases, $\arg[g_4 A_R^*]  $ and $\arg[g_3 A_L^*]$, if $Im[g_4 A_R^*]$ and $Im[g_3 A_L^*]$ have same sign, and if there are no scalar and pseudo scalar interaction. 
Similarly, we can also determine the argument of relative phases, $\arg[g_6 A_R^*] $ and $\arg[g_5 A_L^*]$, if $g_+$ is nonzero.
    
       Even if $\tau\to\mu\gamma$ process is  suppressed by a factor of 100 or more than $\tau\to 3\mu$ process, we may still estimate the branching ratio of  $\tau\to\mu\gamma$ from these interference: 
       \begin{eqnarray}
\frac{\mathrm{Br}(\tau\to 3\mu \ \ \mathrm{interference}) }{\mathrm{Br}(\tau\to\mu\gamma) } 
\sim 
\sqrt{\alpha}\left|\frac{ g_3+g_4+g_5+g_6}{A_R+A_L}\right|
\gg 1,
\end{eqnarray} 
 where $\alpha$ is the fine  structure constant.
In concrete terms, for instance, even if $|eA_L|^2$ is too small to determine directly, we may be still able to get a lower bound $(d_++d_-)^2/\{2(a_+-a_-)\}$ as explained in subsection \ref{a<o}.

\section*{Acknowledgment}

A.M. is supported in part by Grants-in-Aid for Scientific Research from the Ministry of Education, Culture, Sports, Science and Technology of Japan.
The authors thank Tadashi Yoshikawa for his very helpful comments.


\appendix

\section{Fiertz Transformation}

 In Eq. (\ref{lag4f}), the terms 
 \begin{align} \begin{split}
(\bar{\tau}_R \mu_L)&(\bar{\mu}_L  \mu_R),
\\
(\bar{\tau}_L \mu_{R})&(\bar{\mu}_R  \mu_L),
\\
(\bar{\tau}_R \sigma_{\alpha \beta} \mu_L)&(\bar{\mu}_R \sigma^{\alpha \beta}  \mu_L),
\\
(\bar{\tau}_L \sigma_{\alpha \beta} \mu_R)&(\bar{\mu}_L \sigma^{\alpha \beta}  \mu_R),
\\
(\bar{\tau}_L \sigma_{\alpha \beta} \mu_R)&(\bar{\mu}_R \sigma^{\alpha \beta}  \mu_L),
\\
(\bar{\tau}_R \sigma_{\alpha \beta} \mu_L)&(\bar{\mu}_L \sigma^{\alpha \beta}  \mu_R),
\end{split} \end{align}
which are naively assumed particularly don't  appear.
The reasons are as follows.

First, using the equations
 \begin{align} \begin{split}
 &(\bar{\psi}_{1R}\sigma^{\mu\nu}\psi_{2L})(\bar{\psi}_{3R}\sigma_{\mu\nu}\psi_{4L})
 \\&=-6(\bar{\psi}_{1R} \psi_{4L})(\bar{\psi}_{3R} \psi_{2L})
 +\frac{1}{2}(\bar{\psi}_{1R}\sigma^{\mu\nu}\psi_{4L})(\bar{\psi}_{3R}\sigma_{\mu\nu}\psi_{2L})
 \end{split} \end{align}
 and
  \begin{align} \begin{split}
 &(\bar{\psi}_{1R}\sigma^{\mu\nu}\psi_{4L})(\bar{\psi}_{3R}\sigma_{\mu\nu}\psi_{2L})
 \\&=-6(\bar{\psi}_{1R} \psi_{2L})(\bar{\psi}_{3R} \psi_{4L})
 +\frac{1}{2}(\bar{\psi}_{1R}\sigma^{\mu\nu}\psi_{2L})(\bar{\psi}_{3R}\sigma_{\mu\nu}\psi_{4L}),
 \end{split} \end{align}
 where the  $\psi_{iL}=(1-\gamma_5)\psi_i/2$,  $\psi_{iR}=(1+\gamma_5)\psi_i/2$, $\bar \psi_{iL}=\bar\psi_i(1+\gamma_5)/2$,  $\bar\psi_{iR}=\bar\psi_i(1-\gamma_5)/2$ and $\psi_i$ where $i=\{1,2,3,4\}$ are the Dirac spinors.
   Then,
\begin{align} \begin{split}
 &(\bar{\psi}_{1R}\sigma^{\mu\nu}\psi_{2L})(\bar{\psi}_{3R}\sigma_{\mu\nu}\psi_{4L})
 -(\bar{\psi}_{1R}\sigma^{\mu\nu}\psi_{4L})(\bar{\psi}_{3R}\sigma_{\mu\nu}\psi_{2L})
 \\&=-4(\bar{\psi}_{1R} \psi_{4L})(\bar{\psi}_{3R} \psi_{2L})+4(\bar{\psi}_{1R} \psi_{2L})(\bar{\psi}_{3R} \psi_{4L})
 \end{split} \end{align}
and also exchanging L and R, 
 \begin{align} \begin{split}
 &(\bar{\psi}_{1L}\sigma^{\mu\nu}\psi_{2R})(\bar{\psi}_{3L}\sigma_{\mu\nu}\psi_{4R})
 -(\bar{\psi}_{1L}\sigma^{\mu\nu}\psi_{4R})(\bar{\psi}_{3L}\sigma_{\mu\nu}\psi_{2R})
 \\&=-4(\bar{\psi}_{1L} \psi_{4R})(\bar{\psi}_{3L} \psi_{2R})
     +4(\bar{\psi}_{1L} \psi_{4R})(\bar{\psi}_{3L} \psi_{2R}).
 \end{split} \end{align}

Next, 
 \begin{align} \begin{split}
 (\bar{\psi}_{1R}\sigma^{\mu\nu}\psi_{2L})(\bar{\psi}_{3L}\sigma_{\mu\nu}\psi_{4R})
 =\sum_i C_i (\bar{\psi}_{1R}\Gamma_i\psi_{4R})(\bar{\psi}_{3L}\Gamma_i\psi_{2L})=0
 \end{split} \end{align}
where $\Gamma_i=\{1,\gamma_5,\sigma^{\mu\nu}\}$ and $C_i$ are some coefficients.

Finally,
\begin{align} \begin{split}
 (\bar{\psi}_{1L} \psi_{2R})(\bar{\psi}_{3R} \psi_{4L})
=-2(\bar{\psi}_{1L} \gamma^\mu \psi_{4R})(\bar{\psi}_{3L} \gamma_\mu \psi_{2R}).
\end{split} \end{align}
Then,
 \begin{align} \begin{split}
 &(\bar{\psi}_{1L} \psi_{2R})(\bar{\psi}_{3R} \psi_{4L})
 -(\bar{\psi}_{1L} \psi_{4R})(\bar{\psi}_{3R} \psi_{2L})
\\&=2(\bar{\psi}_{1L} \gamma^\mu \psi_{2R})(\bar{\psi}_{3L} \gamma_\mu \psi_{4R})
   -2(\bar{\psi}_{1L} \gamma^\mu \psi_{4R})(\bar{\psi}_{3L} \gamma_\mu \psi_{2R})
\end{split} \end{align}
and also  
 \begin{align} \begin{split}
 &(\bar{\psi}_{1R} \psi_{2L})(\bar{\psi}_{3L} \psi_{4R})
 -(\bar{\psi}_{1R} \psi_{4L})(\bar{\psi}_{3L} \psi_{2R})
\\&=2(\bar{\psi}_{1R} \gamma^\mu \psi_{2L})(\bar{\psi}_{3R} \gamma_\mu \psi_{4L})
   -2(\bar{\psi}_{1R} \gamma^\mu \psi_{4L})(\bar{\psi}_{3R} \gamma_\mu \psi_{2L}).
\end{split} \end{align}
So it is proved that Eq. (\ref{lag4f}) is the general form of 4-Fermi type interactions.

\section{Parts of Master Formula}

\subsection{production cross section of $\tau$ pair with the polarizations}\label{sec e+e-to taupair}
Defining $s^\pm$ as $\tau^\pm$ polarization vector in frame 3 and 2, respectively, the differential cross section for the process  $e^+ e^- \to \gamma^* \to \tau^+(s^+) \tau^-(s^-) $ in the center of mass frame, frame1 is \cite{Sanda}:
\begin{eqnarray}\label{e+e-to taupair}
&&\frac{d\sigma\bigl(e^+ e^- \to \tau^+(s^+) \tau^-(s^-)\bigr)}{d\Omega} \nonumber
\\&=&
\frac{ \alpha^2 \beta }{4 q^2}\biggl[\bigl(1+\cos^2\eta +\frac{\sin ^2 \eta}{\gamma^2}\bigr) +\bigl(1+\cos^2\eta -\frac{\sin ^2 \eta}{\gamma^2}\bigr) s_{z}^+ s_{z}^- 
-\beta^2 \sin ^2 \eta \  s_{y}^+ s_{y}^- \nonumber  \\
&&\hspace{3em} +\bigl(1+\frac{1}{\gamma^2}\bigr)\sin^2\eta \ s_{x}^+ s_{x}^-
-\frac{\sin 2\eta}{\gamma}(s_{z}^+s_{x}^- + s_{x}^+  s_{z}^-)\biggr], 
\end{eqnarray}
where $\alpha \simeq 1/137$ is the  fine structure constant, $\beta=|\mathbf{p}_{\tau}'|/E$, $\gamma=(1-\beta^2)^{-1/2}=E/m_\tau$, E is the energy of $e^+$ or $e^-$ in the initial state and $|\mathbf{p}_{\tau}'|$ is the absolute value of momentum of   $\tau^+$.
 As described in Fig. \ref{fig:frame123.eps}, $\eta$ is the angle between the momenta of $e^+$ in the initial state and $\tau^+$.
$q^2=(p_{e^+}'+p_{e^-}')^2$, where $p_{e^+}'$ and $p_{e^-}'$ are the momenta of $e^+$ and $e^-$ in the initial state, respectively.
$\Omega$ is the solid angle for the $\tau^+$ momentum.
We note that the quantities, $E$, $|\mathbf{p}_{\tau}'|$, $p_{e^\pm}'$ and $\Omega$ are defined in frame 1.

\subsection{differential branching ratio for $\tau^-$ decay }\label{sec tau^-decay}
The differential Branching ratio for the process $\tau^- \to \nu_\tau+a+\mathrm{anything}$ in the rest frame of $\tau^-$ is \cite{Sanda-2};
\begin{eqnarray}\label{tau^-decay}
&&\frac{dBr(\tau^-(s^-) \to \nu_\tau+a+\mathrm{anything})}{d^3k_{a}}  \nonumber
\\&=&Br(\tau^- \to \nu_\tau+a+\mathrm{anything})\frac{2}{\pi m_\tau^3 \lambda_a}\left[ G_1^a(y_a)-  \mathbf{s}^- \cdot \hat {\mathbf{k}}_a  G_2^a(y_a) \right], \nonumber  \\
\end{eqnarray}
where $G_1^a(y_a)$ and $G_2^a(y_a)$ are the functions of $y_a$ defined in each $a$.
 These are written in the table 1 of Ref. \cite{Sanda-2}. 
Here, 
\begin{eqnarray}
\lambda_a=\int dy_a y_a^2 G_1^a(y_a),
\end{eqnarray}
\begin{eqnarray}
y_a=\frac{2E_{a}}{m_\tau}, 
\end{eqnarray}
$E_{a}$ is the energy of $a$,
$\hat {\mathbf{k}}_a=  \mathbf{k}_a  /|\mathbf{k}_a|$ and $k_a$ is the momentum of the particle $a$.

We note here that physical vector quantities which we treat in this process are only $ \mathbf{s}^- $ and $ \hat {\mathbf{k}}_a $.
The only scalar made by these vector quantities is $ \mathbf{s}^- \cdot \hat {\mathbf{k}}_a $
So, we can explain the differential branching ratio, Eq. (\ref{tau^-decay}) by only two terms which are proportional to $G_1^a(y_a)$ and $  \mathbf{s}^- \cdot \hat {\mathbf{k}}_a  G_2^a(y_a)$, respectively.

\subsection{narrow width approximation}\label{narrow width app}
The narrow width approximation is
\begin{eqnarray}
\frac{1}{|p^2-(m-i\Gamma/2)|^2}\simeq \frac{\pi}{m\Gamma}\delta(p^2-m^2), 
 \ \ \mathrm{where} \ \ \ \frac{\Gamma}{m}\ll 1.
\end{eqnarray}

\subsection{the total Branching ratio }
In Eq. (\ref{sigma-non-pol}), if $\sqrt{q^2}=m_{\Upsilon(4s)}$ which is the Upsilon $4S$ mass, the differential cross section in the center of mass frame, frame 1 is
\begin{align} \begin{split}
&\frac{d\sigma}
{       dx_1 dx_2   } 
  \\
&\hspace{1em}=
  G_0(x_1,x_2)  \frac{64 \pi \alpha^2   }{m_{\Upsilon(4s)}^2    }  
\left(1  +  \frac{ 2 m_\tau^2}{m_{\Upsilon(4s)}^2 } \right)
\sqrt{1-\frac{ 4 m_\tau^2 }{m_{\Upsilon(4s)}^2}}
\\
& \hspace{2em} \times
  Br(\tau \to \mu \nu \bar\nu) Br\bigl(\tau^- \to \nu_\tau+a +\mathrm{anything}\bigr).
\end{split} \end{align}
Then, the total cross section is 
\begin{align} \begin{split}
\sigma
&=\in {0} {1 }   \in {1-\frac{x_3}{2}} {1-(\frac{4\delta}{3})^2}
\frac{d\sigma}{       dx_1 dx_3   }  dx_1 dx_3  =
  \frac{64 \pi \alpha^2   }{3 m_{\Upsilon(4s)}^2    }  
\left(1  +  \frac{ 2 m_\tau^2}{m_{\Upsilon(4s)}^2 } \right)
\sqrt{1-\frac{ 4 m_\tau^2 }{m_{\Upsilon(4s)}^2}}
\\
& \hspace{2em} \times
  Br(\tau \to \mu \nu \bar\nu) Br\bigl(\tau^- \to \nu_\tau+a +\mathrm{anything}\bigr)
\\&  \times  \Biggl[ \frac{1}{2}a_+ + \frac{1}{4}b_+
+4d_+ +2e_+
+\frac{2}{3}\Bigl(24\log\bigl[\frac{3}{4\delta}\bigr]-13\Bigr)c_+ \Biggr]
 \end{split} \end{align}
where
\begin{eqnarray*}
\frac{2}{3}\Bigl(24\log\bigl[\frac{3}{4\delta}\bigr]-13\Bigr)
\simeq 20.8
\end{eqnarray*}
since $\delta=2m_\mu/m_\tau$. 
%
%
%
Here, we derive the total cross section in center of mass frame of initial electron and positron since the total cross section is Lorentz invariant under the boost for the beam direction. 
So, this expression is able to apply to the B Factory experiment.

\section{G's}\label{Appendix C}

  Using the width $\Gamma(\tau\to \mu\nu\bar\nu)=m_\tau^5 G_F^2/(192 \pi^3)$,
 the differential branching ratio concerning the polarization of $\tau$ is 
\begin{align} \begin{split}
\frac{d\mathrm{Br}\bigl(\tau^+(s^+)\to \mu_1 \mu_2 \mu_3\bigr)}{dx_1 dx_2   d\Omega_\tau d\psi} 
  =\frac{3}{2\pi^2} \mathrm{Br}(\tau\to \mu \nu\bar\nu)
\Bigl[  G_0(x_1,x_2)+  \mathbf{s}^+ \cdot \mathbf{P}_i G_i^s(x_1,x_2)  \Bigr] \label{full dB}.
\end{split} \end{align} 
$G_0(x_1,x_2)$ and $G_i^s(x_1,x_2)$ are as follows.

\subsection{ $G_0(x_1,x_2)$}

\begin{align} \begin{split}
G_0(x_1,x_2)
=
   &4a_+
      (2-x_1-x_2 )  (x_1+x_2-1 ) 
   + 
    b_+  ( (1-x_1 ) x_1+ (1-x_2 ) x_2 )  
    \\&+4c_+\frac{
      (1-x_1)  (  x_1^2+(1-x_1)^2 )
     +(1-x_2)  (x_2^2 +(1-  x_2)^2 )  }{ (1-x_1 )  (1-x_2 )}
   \\&  +16 d_+  (x_1+x_2-1 )
   +4  e_+  (2-x_1-x_2 )  
\end{split} \end{align}

\subsection{$G_i^s(x_1,x_2)$}

\begin{align} \begin{split}
G_1^s(x_1,x_2)
=
     &-4 a_-
   x_1  (x_1+x_2-1 )  
   +  
    b_- x_1 (1-x_1 )   
   \\&+4c_-\frac{
     x_1  \bigl(2x_1  (1-x_1 )+2( x_1+ x_2-1 )-1\bigr) }{ (1-x_1 )  (1-x_2 )}
\\&-4 d_-\frac{  x_1  (x_1+x_2-1 )
    (-x_2^2+x_1  (x_2-2 )+2 )  }{ (1-x_1 )  (1-x_2 )}
    \\&-2e_-\frac{
      x_1    \bigl( (2-x_2 ) (1-x_1)^2+ x_2(1-x_2 )^2 \bigr)   }{ (1-x_1 )  (1-x_2 )}
\end{split} \end{align}

\begin{align} \begin{split}
G_2^s(x_1,x_2)
=G_1^s(x_2,x_1)
\end{split} \end{align}

\begin{align} \begin{split}
G_3^s(x_1,x_2)
=
\Bigl[
  -2f_+ (x_1+x_2-1 )  -g_+ (x_1+x_2-2 ) 
   \Bigr]
    \frac{2
          x_1  (x_1-x_2 ) x_2 }{ (x_1-1 )  (x_2-1 )}
\end{split} \end{align}

\section{derivations of relations, $a_+c_+-d_+^2$ and $b_+c_+-e_+^2$}\label{6.1}

Here, we introduce a convenient formulae $a_+c_+-d_+^2$ and $b_+c_+-e_+^2$ to use in sections \ref{phys of 1}, \ref{phys of 3}, \ref{E 1,3} and \ref{physics implication}.
Using this formula, we can determine more about current structure.

Using the relations
\begin{align} \begin{split}
&a_+ c_+ - |g_3 eA_L^*+g_4 eA_R^*|^2 
\\&=\frac{1}{16}(r_1^2+r_2^2)(r_L^2+r_R^2)+(r_3^2+r_4^2)(r_L^2+r_R^2)\\
&\quad -r_3^2 r_L^2-r_4^2 r_R^2 -2r_3 r_L r_4 r_R \cos(\theta_3-\theta_L-\theta_4+\theta_R)\\
&=r_3^2r_R^2-2r_3 r_R r_4 r_L \cos(\theta_3+\theta_R-\theta_4-\theta_L)+r_4^2 r_L^2+\frac{1}{16}(r_1^2+r_2^2)(r_L^2+r_R^2)\\
&=|r_3e^{i\theta_3} r_Re^{i\theta_R}- r_4e^{i\theta_4} r_Le^{i\theta_L}|^2+\frac{1}{16}(r_1^2+r_2^2)(r_L^2+r_R^2)\\
&=|g_3 eA_R-g_4 eA_L|^2+\frac{1}{16}(r_1^2+r_2^2)(r_L^2+r_R^2)
\end{split} \end{align}
and
\begin{align} \begin{split}
& |g_3 eA_L^*+g_4 eA_R^*|^2 - d_+^2\\
 &=r_3^2 r_L^2+r_4^2 r_R^2+2r_3 r_L r_4 r_R \cos(\theta_3-\theta_L-\theta_4+\theta_R)
 \\&\quad-r_3^2 r_L^2\cos^2(\theta_3-\theta_L)-2r_3 r_L r_4 r_R\cos(\theta_3-\theta_L)\cos(\theta_4-\theta_R)- r_4^2 r_R^2\cos^2(\theta_4-\theta_R)
 \\&=r_3^2 r_L^2\sin^2(\theta_3-\theta_L)
 +2r_3 r_L r_4 r_R\sin(\theta_3-\theta_L)\sin(\theta_4-\theta_R)
 +r_4^2 r_R^2\sin^2(\theta_4-\theta_R)
 \\&=\bigl\{r_3 r_L\sin(\theta_3-\theta_L)
 +r_4 r_R\sin(\theta_4-\theta_R)\bigr\}^2
 \\&=\bigl(Im[g_3eA_L^*]+Im[g_4eA_R^*]\bigr)^2,
\end{split} \end{align}
we give a useful relation
\begin{align} \begin{split}\label{Eq h}
&a_+ c_+ - d_+^2 
\\
&=\frac{1}{16}(r_1^2+r_2^2)(r_L^2+r_R^2)+|r_3e^{i\theta_3} r_Re^{i\theta_R}- r_4e^{i\theta_4} r_Le^{i\theta_L}|^2
\\&\hspace{9em}+\bigl\{r_3 r_L\sin(\theta_3-\theta_L)
 +r_4 r_R\sin(\theta_4-\theta_R)\bigr\}^2\\
&=\frac{1}{16}(|g_1|^2+|g_2|^2)(|eA_L|^2+|eA_R|^2)+|g_3 eA_R-g_4 eA_L|^2\\&\hspace{10em}+\bigl(Im[g_3eA_L^*]+Im[g_4eA_R^*]\bigr)^2\ge 0.
\end{split} \end{align}
 
In this formula, there are three terms.
Each of them has zero or positive value.
First term $(|g_1|^2+|g_2|^2)(|eA_L|^2+|eA_R|^2)/16$ has the information about scalar and pseudo scalar currents.
Second term $|g_3 eA_R-g_4 eA_L|^2$ has information about parity symmetry.
Third term $\bigl(Im[g_3eA_L^*]+Im[g_4eA_R^*]\bigr)^2$ has information about CP symmetry.

If $a_+=0$, then $r_1=r_2=r_3=r_4=0$ and also $d_+=0$.  
Similarly, if $c_+=0$, then $r_R=r_L=0$ and also $d_+=0$. 
In these cases, we can't use interference effect.  
So, we consider only the case 
\begin{align} \begin{split}\label{3}
a_+& \not=0 \\
 c_+& \not=0
 \end{split} \end{align}
when we use this formula.

Similarly, 
\begin{align} \begin{split}\label{Eq h2}
&b_+ c_+ - e_+^2 
\\
&=|r_5e^{i\theta_5} r_Re^{i\theta_R}- r_6e^{i\theta_6} r_Le^{i\theta_L}|^2+\bigl\{r_5 r_L\sin(\theta_5-\theta_L)
 +r_6 r_R\sin(\theta_6-\theta_R)\bigr\}^2\\
&=|g_5 eA_R-g_6 eA_L|^2+\bigl(Im[g_5eA_L^*]+Im[g_6eA_R^*]\bigr)^2\ge 0.
\end{split} \end{align}
 
In this formula, there are two terms.
Each of them has zero or positive value.
First term $|g_5 eA_R-g_6 eA_L|^2$ has information about parity symmetry.
Second term $\bigl(Im[g_5eA_L^*]+Im[g_6eA_R^*]\bigr)^2$ has information about CP symmetry.

If $b_+=0$, then $r_5=r_6=0$ and also $e_+=0$.  
Similarly, if $c_+=0$, then $r_R=r_L=0$ and also $e_+=0$. 
In these cases, we can't use interference effect.  
So, we consider only the case 
\begin{align} \begin{split}\label{32}
b_+& \not=0 \\
 c_+& \not=0
 \end{split} \end{align}
when we use this formula.

\section{C, P, T and CP Transformation }\label{CPT}
We define charge (C), parity (P), time reversal (T) and CP transformations.

In Lagrangian, C transformation is
\begin{align}
g_1 &\longleftrightarrow  g_2^*&\hspace{-2em}
g_2 &\longleftrightarrow  g_1^*\nonumber\\
g_3 &\longleftrightarrow  g_4^*&\hspace{-2em}
g_4 &\longleftrightarrow  g_3^*\nonumber\\
g_5 &\longleftrightarrow  g_6^*&\hspace{-2em}
g_6 &\longleftrightarrow  g_5^*\\
A_R &\longleftrightarrow  A_L^*&\hspace{-2em}
A_L &\longleftrightarrow  A_R^*,\nonumber
\end{align}
P transformation is
\begin{align}
g_1 &\longleftrightarrow  g_2&\hspace{-2em}
g_1^* &\longleftrightarrow  g_2^*\nonumber\\
g_3 &\longleftrightarrow  g_4&\hspace{-2em}
g_3^* &\longleftrightarrow  g_4^*\nonumber\\
g_5 &\longleftrightarrow  g_6&\hspace{-2em}
g_5^* &\longleftrightarrow  g_6^*\\
A_R &\longleftrightarrow  A_L&\hspace{-2em}
A_R^* &\longleftrightarrow  A_L^*\nonumber
\end{align}
and 
T and CP transformations are
\begin{align}
g_1 &\longleftrightarrow  g_1^*&\hspace{-2em}
g_2 &\longleftrightarrow  g_2^*\nonumber\\
g_3 &\longleftrightarrow  g_3^*&\hspace{-2em}
g_4 &\longleftrightarrow  g_4^*\nonumber\\
g_5 &\longleftrightarrow  g_5^*&\hspace{-2em}
g_6 &\longleftrightarrow  g_6^*\\
A_R &\longleftrightarrow  A_R^*&\hspace{-2em}
A_L &\longleftrightarrow  A_L^*.\nonumber
\end{align}

\section{Observables in $e^+ e^- $ Center of Mass Frame}

 The relation
\begin{align} \begin{split}
\frac{d\sigma}{d\alpha_1 d\alpha_2 \cdots d\alpha_n}=\int \frac{d\sigma}{d\beta_1d\beta_2\cdots,d\beta_m}
\prod _{i=1}^n \delta\bigl(\alpha_i-\alpha_i(\beta_1,\beta_2,\cdots,\beta_m)\bigr)d\beta_1d\beta_2\cdots d\beta_m
\end{split} \end{align}
is useful for converting the variables from $\beta_1,\beta_2,\cdots,\beta_m $ to $\alpha_1,\alpha_2 ,\cdots ,\alpha_n $ \cite{Sanda}.

To write it briefly, we define $s\eta$, $c\eta$, $ s\phi$, $ c\phi$, $ s\theta$, $ c\theta$, $ s\psi$, $ c\psi$, $ s\xi$, $ c\xi$, $ s\chi$ and $ c\chi  $ as $\sin \eta$, $\cos \eta$, $ \sin \phi$, $ \cos \phi$, $ \sin \theta$, $ \cos \theta$, $ \sin \psi$, $ \cos \psi$, $ \sin \xi$, $ \cos \xi$, $ \sin \chi$ and $ \cos \chi  $, respectively.
Also we define the quantities $p_1',p_2',p_3',k_a'$ as the quantities $p_1,p_2,p_3,k_a$ in frame 1.
First,
\begin{align} \begin{split}
 p_3'&=\left(
  \begin{array}{cccc}
    1   & 0   & 0   & 0   \\
    0   & c\eta   & 0   &  s\eta  \\
    0   & 0   & 1   &  0  \\
    0   & -s\eta   & 0  & c\eta   \\
  \end{array}
\right)
\left(
  \begin{array}{cccc}
    \gamma   &  0  &  0  & \gamma\beta   \\
   0    & 1   & 0   & 0   \\
   0    & 0   & 1   & 0   \\
   \gamma\beta    & 0   & 0   & \gamma   \\
  \end{array}
\right)
\\
&\ \ \ \ \times
\left(
  \begin{array}{cccc}
    1   &  0  & 0   & 0   \\
    0   & c\phi   & -s\phi   &  0  \\
    0   & s\phi   &  c\phi  &  0  \\
    0   &  0  &  0  & 1   \\
  \end{array}
\right)
\left(
  \begin{array}{cccc}
    1   & 0   & 0   & 0   \\
    0   & c\theta   & 0   & s\theta  \\
    0   & 0   & 1   & 0   \\
    0   & -s\theta   & 0  & c\theta   \\
  \end{array}
\right)
\left(
  \begin{array}{c}
    E_3   \\
    0   \\
    0   \\
    E_3   \\
  \end{array}
\right)
\\
&=E_3
\left(
  \begin{array}{cccc}
    \gamma   &  0  &  0  & \gamma\beta   \\
   s\eta \gamma\beta     & c\eta   & 0   & s\eta \gamma   \\
   0    & 0   & 1   & 0   \\
   c\eta \gamma\beta    & -s\eta  & 0   &c\eta \gamma   \\
  \end{array}
\right)
\left(
  \begin{array}{c}
    1   \\
   s\theta c\phi   \\
   s\theta s\phi    \\
    c\theta   \\
  \end{array}
\right)
\\
&=E_3\left(
  \begin{array}{cccc}
    \gamma(1+\beta c\theta)        \\
     c\eta s\theta c\phi+s\eta\gamma(\beta+c\theta)   \\
     s\theta s\phi    \\
    -s\eta s\theta c\phi+c\eta\gamma(\beta+c\theta)       \\
  \end{array}
\right).
\end{split} \end{align}
Here,
$\theta$ and $\phi$ which are defined in Fig. \ref{fig:frame3thetaphi.eps} represent the direction of $\mathbf p_3$ in frame 3.
 $\eta$ represents the direction of $\tau^+$ in frame 1.

Next,
\begin{align} \begin{split}
p_1'&=\left(
  \begin{array}{cccc}
    1   & 0   & 0   & 0   \\
    0   & c\eta   & 0   &  s\eta  \\
    0   & 0   & 1   &  0  \\
    0   & -s\eta   & 0  & c\eta   \\
  \end{array}
\right)
\left(
  \begin{array}{cccc}
    \gamma   &  0  &  0  & \gamma\beta   \\
   0    & 1   & 0   & 0   \\
   0    & 0   & 1   & 0   \\
   \gamma\beta    & 0   & 0   & \gamma   \\
  \end{array}
\right)
\left(
  \begin{array}{cccc}
    1   &  0  & 0   & 0   \\
    0   & c\phi   & -s\phi   &  0  \\
    0   & s\phi   &  c\phi  &  0  \\
    0   &  0  &  0  & 1   \\
  \end{array}
\right)
\\&\times 
\left(
  \begin{array}{cccc}
    1   & 0   & 0   & 0   \\
    0   & c\theta   & 0   & s\theta  \\
    0   & 0   & 1   & 0   \\
    0   & -s\theta   & 0  & c\theta   \\
  \end{array}
\right)
\left(
  \begin{array}{cccc}
   1   &  0  & 0   & 0   \\
    0   & c\psi   & -s\psi   &  0  \\
    0   & s\psi   &  c\psi  &  0  \\
    0   &  0  &  0  & 1   \\
  \end{array}
\right)
\left(
  \begin{array}{cccc}
   1   & 0   & 0   & 0   \\
    0   & c\xi   & 0   & s\xi  \\
    0   & 0   & 1   &  0  \\
    0   & -s\xi   & 0  & c\xi   \\
  \end{array}
\right)
\left(
  \begin{array}{c}
    1   \\
    0   \\
    0   \\
    1   \\
  \end{array}
\right)E_1
\\&= 
\left(
  \begin{array}{cccc}
    \gamma   &  0  &  0  & \gamma\beta   \\
   s\eta \gamma\beta     & c\eta   & 0   & s\eta \gamma   \\
   0    & 0   & 1   & 0   \\
   c\eta \gamma\beta    & -s\eta  & 0   &c\eta \gamma   \\
  \end{array}
\right)
\left(
  \begin{array}{c}
    1   \\
 c\phi( c\theta c\psi  s\xi + s\theta c\xi)   -s\phi( s\psi  s\xi ) \\
   s\phi( c\theta c\psi  s\xi + s\theta c\xi)  + c\phi( s\psi  s\xi )  \\
   -s\theta c\psi  s\xi +  c\theta  c\xi  \\
  \end{array}
\right)E_1, 
\end{split} \end{align}
where $\psi$ is the angle between $\mathbf{p}_2$-$\mathbf{p}_3$ plane and $\mathbf{p}_3$-$z$ plane as represented in Fig. \ref{fig:frame3p1p2.eps}. 
Similarly,
\begin{align} \begin{split}
 p_2'&=\left(
  \begin{array}{cccc}
    \gamma   &  0  &  0  & \gamma\beta   \\
   s\eta \gamma\beta     & c\eta   & 0   & s\eta \gamma   \\
   0    & 0   & 1   & 0   \\
   c\eta \gamma\beta    & -s\eta  & 0   &c\eta \gamma   \\
  \end{array}
\right)
\left(
  \begin{array}{c}
    1   \\
 c\phi( c\theta c\psi  s\chi + s\theta c\chi)   -s\phi( s\psi  s\chi ) \\
   s\phi( c\theta c\psi  s\chi + s\theta c\chi)  + c\phi( s\psi  s\chi )  \\
   -s\theta c\psi  s\chi +  c\theta  c\chi  \\
  \end{array}
\right)
E_2.
\end{split} \end{align}

Here, as Eq. (\ref{Eq d}),
\begin{align} \begin{split}
\cos\xi &=1-2\frac{1-x_2}{x_1 x_3}=\hat{\mathbf{p}}_3 \cdot \hat{\mathbf{p}}_1 \\
\sin\xi &=\sqrt{1-\cos^2\xi}
\end{split} \end{align}
and
\begin{align} \begin{split}
\cos\chi&=1-2\frac{1-x_1}{x_2 x_3}=\hat{\mathbf{p}}_3 \cdot \hat{\mathbf{p}}_2 \\
\sin\chi&=-\sqrt{1-\cos^2\chi},
\end{split} \end{align}
where $\sin\chi \le 0$ since $\pi \le \chi \le 2\pi$.
 
Finally, we use the 3 ways to explain $k_a'$ for the benefit of the simplicity of the phase space integral.
The coordinates are as Figs. \ref{fig:frame2thetaphi-z.eps}, \ref{fig:frame2thetaphi-x.eps} and \ref{fig:frame2thetaphi-y.eps}, respectively.
\begin{align} \begin{split}
k_a'&=
\left(
  \begin{array}{cccc}
    1   & 0   & 0   & 0   \\
    0   & c\eta   & 0   &  s\eta  \\
    0   & 0   & 1   &  0  \\
    0   & -s\eta   & 0  & c\eta   \\
  \end{array}
\right)
\left(
  \begin{array}{cccc}
    \gamma   &  0  &  0  & -\gamma\beta   \\
   0    & 1   & 0   & 0   \\
   0    & 0   & 1   & 0   \\
  - \gamma\beta    & 0   & 0   & \gamma   \\
  \end{array}
\right)
\left(
  \begin{array}{c}
    E_a   \\
   |\mathbf{k}_a| \sin\theta_{az} \cos \phi_{az}   \\
   |\mathbf{k}_a| \sin\theta_{az} \sin \phi_{az}   \\
   |\mathbf{k}_a| \cos\theta_{az}    \\
  \end{array}
\right)
\\&=
\left(
  \begin{array}{cccc}
      \gamma   &  0  &  0  &- \gamma\beta   \\
      -s\eta \gamma\beta &  c\eta  &  0  & s\eta \gamma   \\
      0 & 0   &  1  &  0  \\
      -c\eta \gamma\beta & -s\eta   &  0  &  c\eta \gamma  \\
  \end{array}
\right)
\left(
  \begin{array}{c}
    E_a   \\
   |\mathbf{k}_a| \sin\theta_{az} \cos \phi_{az}   \\
   |\mathbf{k}_a| \sin\theta_{az} \sin \phi_{az}   \\
   |\mathbf{k}_a| \cos\theta_{az}    \\
  \end{array}
\right).
\end{split} \end{align}

\begin{align} \begin{split}
k_a'&=
\left(
  \begin{array}{cccc}
      \gamma   &  0  &  0  &- \gamma\beta   \\
      -s\eta \gamma\beta &  c\eta  &  0  & s\eta \gamma   \\
      0 & 0   &  1  &  0  \\
      -c\eta \gamma\beta & -s\eta   &  0  &  c\eta \gamma  \\
  \end{array}
\right)
\left(
  \begin{array}{c}
    E_a   \\
   |\mathbf{k}_a| \cos\theta_{ax}   \\
   |\mathbf{k}_a| \sin\theta_{ax} \cos \phi_{ax}   \\
   |\mathbf{k}_a| \sin\theta_{ax} \sin \phi_{ax}    \\
  \end{array}
\right).
\end{split} \end{align}

\begin{align} \begin{split}
k_a'&=
\left(
  \begin{array}{cccc}
      \gamma   &  0  &  0  & -\gamma\beta   \\
      -s\eta \gamma\beta &  c\eta  &  0  & s\eta \gamma   \\
      0 & 0   &  1  &  0  \\
      -c\eta \gamma\beta & -s\eta   &  0  &  c\eta \gamma  \\
  \end{array}
\right)
\left(
  \begin{array}{c}
    E_a   \\
   |\mathbf{k}_a| \sin\theta_{ay} \sin \phi_{ay}   \\
   |\mathbf{k}_a| \cos\theta_{ay}   \\
   |\mathbf{k}_a| \sin\theta_{ay} \cos \phi_{ay}    \\
  \end{array}
\right).
\end{split} \end{align}

 \end{document}